\documentclass[sigconf, nonacm]{acmart}

\usepackage[utf8]{inputenc}
\usepackage{multirow}
\usepackage{longtable}
\usepackage{graphicx}
\usepackage{adjustbox}
\usepackage{subcaption}
\usepackage{soul}
\usepackage{comment}
\usepackage{xspace}
\usepackage{cleveref}

\usepackage{colortbl}
\usepackage{url}

\usepackage{cleveref}
\usepackage{hyperref}

\usepackage{booktabs}
\usepackage{tabularx}

\usepackage{makecell}

\usepackage[most]{tcolorbox}
\newtcolorbox{mybox}{
  colback=gray!20, % Background color of the box: gray with 20% black
  colframe=gray!60, % Frame color
  boxrule=0.5pt, % Width of the frame
  arc=4pt, % Radius of the corners
  top=5pt, % Top space within the box
  bottom=5pt, % Bottom space within the box
  left=10pt, % Left space within the box
  right=10pt, % Right space within the box
  boxsep=5pt, % Space between text and frame
  breakable % Allows box to split over pages
}
\lstdefinestyle{htmlcssjs}{
  backgroundcolor=\color{gray!10}, % Background color
  basicstyle=\ttfamily\small, % Set font
  frame=single, % Add a frame around the code
  framesep=10pt, % Frame separation
  breaklines=true, % Wrap lines
  postbreak=\mbox{\textcolor{red}{$\hookrightarrow$}\space}, % Mark wrapped line beginnings
}
\usepackage[most]{tcolorbox}
\newtcbox{\highlight}[1][red]{on line,
  colframe=#1!50!black, colback=#1!10!white, 
  boxrule=0.5pt, arc=4pt, boxsep=0pt, left=1pt, right=1pt, top=2pt, bottom=2pt}
  
\usepackage{tikz}
\newcommand*\circled[1]{\tikz[baseline=(char.base)]{
          \node[shape=circle,draw,inner sep=2pt] (char) {#1};}}
        
\newcommand{\name}{\textsc{CookieGuard}\xspace}

\newcommand{\todo}[1]{}
\renewcommand{\todo}[1]{{\color{red} TODO: {#1}}}
\newcommand{\TODO}[1]{}
\renewcommand{\TODO}[1]{{\color{red} TODO: {#1}}}

\usepackage{upquote}

\usepackage{xcolor}

\usepackage{listings}

\newcommand*\dash{\unskip\kern.16667em---\penalty\exhyphenpenalty
        \hskip.16667em\relax
}

 {\makeatletter
  \gdef\xxxmark{%
    \expandafter\ifx\csname @mpargs\endcsname\relax % in minipage?
      \expandafter\ifx\csname @captype\endcsname\relax % in figure/caption?
        \marginpar{\textcolor{red}{xxx}}% not in a caption or minipage, can use marginpar
      \else
        \textcolor{red}{xxx~}% notice trailing space
      \fi
    \else
      \textcolor{red}{xxx~}% notice trailing space
    \fi}
  \gdef\xxx{\@ifnextchar[\xxx@lab\xxx@nolab}
  \long\gdef\xxx@lab[#1]#2{{\bfseries [\xxxmark \textcolor{red}{#2}
  ---{\scshape #1}]}}
  \long\gdef\xxx@nolab#1{{\bfseries [\xxxmark \textcolor{red}{#1}]}}
 }
 
 {\makeatletter
  \gdef\yyymark{%
    \expandafter\ifx\csname @mpargs\endcsname\relax % in minipage?
      \expandafter\ifx\csname @captype\endcsname\relax % in figure/caption?
        \marginpar{\textcolor{blue}{yyy}}% not in a caption or minipage, can use marginpar
      \else
        \textcolor{blue}{yyy~}% notice trailing space
      \fi
    \else
      \textcolor{blue}{yyy~}% notice trailing space
    \fi}
  \gdef\yyy{\@ifnextchar[\yyy@lab\yyy@nolab}
  \long\gdef\yyy@lab[#1]#2{{\bfseries [\yyymark \textcolor{blue}{#2}
  ---{\scshape #1}]}}
  \long\gdef\yyy@nolab#1{{\bfseries [\yyymark \textcolor{blue}{#1}]}}
 }
 
 {\makeatletter
  \gdef\zzzmark{%
    \expandafter\ifx\csname @mpargs\endcsname\relax % in minipage?
      \expandafter\ifx\csname @captype\endcsname\relax % in figure/caption?
        \marginpar{\textcolor{green}{zzz}}% not in a caption or minipage, can use marginpar
      \else
        \textcolor{green}{zzz~}% notice trailing space
      \fi
    \else
      \textcolor{green}{zzz~}% notice trailing space
    \fi}
  \gdef\zzz{\@ifnextchar[\zzz@lab\yyy@nolab}
  \long\gdef\zzz@lab[#1]#2{{\bfseries [\zzzmark \textcolor{green}{#2}
  ---{\scshape #1}]}}
  \long\gdef\zzz@nolab#1{{\bfseries [\zzzmark \textcolor{green}{#1}]}}
 }

\copyrightyear{2025}
\acmYear{2025}
\setcopyright{cc}
\setcctype{by}
\acmConference[IMC '25]{Proceedings of the 2025 ACM Internet Measurement
Conference}{October 28--31, 2025}{Madison, WI, USA}
\acmBooktitle{Proceedings of the 2025 ACM Internet Measurement Conference
(IMC '25), October 28--31, 2025, Madison, WI, USA}
\acmDOI{10.1145/3730567.3764490}
\acmISBN{979-8-4007-1860-1/2025/10}

\begin{document}

\title{\name: Characterizing and Isolating the First-Party Cookie Jar}

\author{Pouneh Nikkhah Bahrami}
\email{pnikkhah@ucdavis.edu}
\affiliation{%
  \institution{University of California}
  \city{Davis}
  \country{USA}
}

\author{Aurore Fass}
\email{fass@cispa.de}
\affiliation{%
  \institution{CISPA Helmholtz Center for Information Security}
  \city{Saarbr\"{u}cken}
  \country{Germany}
}

\author{Zubair Shafiq}
\email{zubair@ucdavis.edu }
\affiliation{%
  \institution{University of California}
  \city{Davis}
  \country{USA}
}


\begin{abstract}
As third-party cookies are being phased out or restricted by major browsers, first-party cookies are increasingly being used for web tracking. 
Prior work has shown that third-party scripts embedded in the main frame can access and exfiltrate first-party cookies—including those set by other third-party scripts. 
However, existing browser security mechanisms such as the Same-Origin Policy (SOP), Content Security Policy (CSP), and third-party storage partitioning do not prevent cross-domain access to first-party cookies in the main frame. 
While recent studies have begun to highlight this issue, there remains a lack of comprehensive measurement and practical defenses.

In this work, we conduct the first large-scale measurement and analysis of cross-domain access to first-party cookies in the main frame for 20,000 websites. 
We find that 56\% of the websites include third-party scripts that exfiltrate first-party cookies that they did not originally set, and 32\% where such scripts overwrite or delete these first-party cookies.
To mitigate potential confidentiality and integrity risks due to this lack of isolation, we propose \name, a browser-based runtime mechanism to isolate first-party cookies on a per-script-origin basis. 
\name blocks unauthorized cross-domain cookie operations while preserving site functionality, with only 3\% of the tested websites being affected by Single Sign-On (SSO) breakage.
Our work highlights the risks posed by the lack of first-party cookie isolation in the current browser security model and offers a deployable path toward stronger protection.

\end{abstract}



% make the title area
\maketitle

\section{Introduction} 
As third-party cookies are being phased out or restricted by major browsers~\cite{thirdpartyblockinmozilla,thirdpartyblockinedge,thirdpartyblockinsafari,thirdpartyblockinchrome}, trackers are increasingly relying on first-party cookies for tracking~\cite{munir2022cookiegraph}.
One technique, known as \textit{cookie ghost-writing}, allows third-party scripts embedded in the main frame to set cookies that browsers treat as first-party~\cite{chen2021cookieswap,sanchez2021journey}.
Crucially, browsers do not distinguish between genuine first-party cookies that are actually set by the first-party and those set by third-party scripts.

Scripts in the main frame (regardless of their origin) have full access to the first-party cookies in the cookie jar. 
This allows any third-party script in the main frame to read, modify, delete, or exfiltrate any first-party cookie, including those set by the first-party or other third-party scripts~\cite{munir2022cookiegraph,chen2021cookieswap}. 
Existing browser security mechanisms such as Same-Origin Policy (SOP), Content Security Policy (CSP), and partitioned storage (e.g., Safari’s ITP~\cite{safariPartitioning}) do not prevent such cross-domain access to first-party cookies (henceforth, \textit{cross-domain cookie interaction}) in the main frame.\footnote{We define a \textit{cross-domain cookie interaction} as any read, write, delete, or exfiltration attempt by a script targeting a cookie it did not originally set.}
While this issue has recently gained attention~\cite{sanchez2021journey,chen2021cookieswap}, prior work lacks both large-scale measurement and effective mitigation.

We address the following research questions in this work:

\begin{enumerate}
    \item How prevalent are cross-domain cookie interactions?
    \item Can we enforce domain-level isolation to mitigate this issue without causing website breakage?
\end{enumerate}

To answer these research questions, we conduct a large-scale measurement and analysis of first-party cookies on 20,000 websites. 
We find that 56\% of the websites include third-party scripts that exfiltrate cookies they did not set, and 32\% where such scripts overwrite or delete these first-party cookies.
To mitigate this issue, we present \name, a prototype browser-based runtime mechanism to isolate first-party cookies on a per-script-domain basis. 
Implemented as a browser extension, \name intercepts calls to \texttt{document.cookie} and \texttt{cookieStore}, allowing each script to access only the cookies it originally set. 
By enforcing policies based on execution context rather than script content, \name is robust against code obfuscation since a script's domain can still be identified regardless of code complexity. 
Moreover, unlike blocklist-based defenses \cite{EasyList, EasyPrivacy} that struggle against domain or URL manipulation \cite{storey2017future}, \name does not rely on enumerating tracker domains; it enforces isolation across \textit{all} domains by design, providing stronger and more durable  protection.
Our evaluation shows that \name blocks all cross-domain cookie interactions while preserving functionality on most websites. 
It introduces a modest average page load overhead of 0.3 seconds and breaks Single Sign-On on 11\% of the websites.
Domain-level policies and organizational groupings help reduce this breakage to only 3\%.

Our key contributions are as follows.
\begin{enumerate}
    \item \textbf{Comprehensive measurement}. We conduct the first large-scale quantification of cross-domain cookie interactions (specifically overwrites and deletions) across 20,000 websites, going beyond prior work focused primarily on exfiltration \cite{chen2021cookieswap, fouad2018missed, sanchez2021journey, munir2022cookiegraph, demir2022towards, dimova2021cname, ren2021analysis, fouad2022my}. We further analyze the potential intent behind manipulations (e.g., competition vs. collusion) and treat exfiltration as a complementary dimension of our analysis.

    \item \textbf{Attribution and use of first-party cookies}. We introduce domain-level attribution of cookie operations, identifying which scripts set, access, or manipulate cookies. Using this framework, we show how first-party cookies are increasingly used for tracking in the post–third-party cookie era, covering both \texttt{document.cookie} and the newer \texttt{CookieStore} API.

    \item \textbf{Prototype enforcement and evaluation}. We design and implement \name, a prototype browser extension that enforces per-script-domain isolation of first-party cookies. We evaluate \name's effectiveness in mitigating cross-domain exfiltration and manipulation, as well as its performance overhead and website breakage.
\end{enumerate}
While \name mitigates cross-domain cookie interactions, it does not address all security and privacy risks posed by third-party scripts in the main frame. 
More broadly, our findings underscore the need for finer-grained isolation in current browser security models. 
To support reproducibility and encourage further research, we make the source code of \name's browser extension implementation publicly available at \url{https://github.com/pooneh-nb/cookieGuard}.
% April 15
\section{Background} 
This section provides an overview of browser security mechanisms, script inclusion practices, and cookie management, with a focus on how first-party cookies can be accessed and manipulated by scripts in the main frame.
An interested reader is referred to \cite{vekaria2025sok} for further discussion and background  of cookies and online tracking. 

% 1. what is main-frame, shared resources, SOP
\subsection{Security Mechanisms}\label{subsec:security_mechanisms}
Modern browsers enforce several security models to protect users and enforce origin boundaries. We summarize those most relevant to script behavior and cookie access.

\textbf{Same-Origin Policy (SOP).} SOP ~\cite{sameoriginpolicy} restricts interactions between resources from different origins, where an origin is defined by the scheme, host, and port. 
For example, \url{https://www.example.com:443} represents one origin. 
SOP isolates frames with different origins and prevents scripts embedded in frames of different origins from accessing cross-origin resources, helping mitigate attacks like cross-site request forgery. 
However, all scripts (from any domain) embedded in the main frame inherit the first-party origin and can access its resources.

In this paper, we explicitly distinguish between cross-origin and cross-domain interactions. 
The domain (eTLD+1) is part of the origin, which includes the full domain name, scheme, and port. For example, \url{https://example.com:8080} and \url{https://subdomain.example.com:8080} have different origins due to varied host names, despite the same scheme, port, and domain. 
We reserve the term cross-origin for SOP's strict definition, while we use cross-domain to denote interactions between scripts from different eTLD+1 (or domains) but executing within the same main-frame origin. 
This distinction is important because our measurements and defenses in this paper focus on cross-domain interactions that deviate from cross-origin interactions that are targeted by SOP.

% CSP
\textbf{Content Security Policy (CSP).}
CSP~\cite{csp} allows site owners to restrict the sources of executable scripts, images, and styles via HTTP headers or meta tags. 
While CSP allows some control over script inclusion, it does not regulate cookie access or define which scripts may read or modify cookies.

\textbf{Storage partitioning.}
Some browsers have introduced storage partitioning to mitigate cross-site tracking. 
Safari implements partitioned cookies through Intelligent Tracking Prevention (ITP) \cite{safariPartitioning}, while Firefox enforces Total Cookie Protection, isolating all storage (including cookies) on a per-site basis \cite{MozillaPartitioning}. 
Chrome is experimenting with Cookies Having Independent Partitioned State (CHIPS), an opt-in model for cookie partitioning under the Privacy Sandbox project \cite{chromePartitioning}.
Although these mechanisms limit third-party tracking across sites, they do not isolate scripts within the same top-level context—allowing third-party scripts in the main frame to access and manipulate first-party cookies.

\subsection{Inclusion of Scripts}
Scripts can be included either directly—through a \texttt{<script>} tag or inline JavaScript—or indirectly, when a loaded script dynamically inserts other scripts using \texttt{eval}, \texttt{import()}, or DOM APIs.

Scripts executing in the main frame, regardless of their original source, are treated as part of the first-party origin and granted full access to shared resources such as cookies and the DOM. 
SOP restricts cross-origin iframe content, but scripts in the main frame can freely interact with all main-frame resources unless explicitly restricted (e.g., via CSP or sandboxing).

\subsection{Cookie Management in Browsers}
\label{sec:background_cookie_management}
Cookies are created and accessed either via HTTP headers or client-side JavaScript.

% http cookie
\textbf{HTTP cookies.} HTTP response header are classified as first-party if they originate from the same domain as the visited site.
Otherwise, they are considered third-party. 
These cookies are automatically attached to future HTTP requests matching the domain (also depending on path and other cookie attributes). 
Cookies marked as \texttt{HttpOnly} are inaccessible to JavaScript, mitigating risks such as Cross-Site Scripting (XSS).

% js cookie
\textbf{Script cookies.} 
Scripts in the main frame can also create and access cookies using the \texttt{document.cookie} or \texttt{cookieStore} API, which provides a string interface to the browser's cookie jar. 
Although SOP governs access, the executing context is considered the visited site’s origin. 
Thus, third-party scripts included in the main frame can create and read first-party cookies, regardless of their actual domain. 
These cookies are effectively indistinguishable from those set by actual first-party scripts.

\textit{document.cookie:}
The \texttt{document.cookie}  provides an interface for reading and writing cookies.
Any script in the main frame can access all first-party cookies for that origin, except those marked as \texttt{HttpOnly}.
Although SOP restricts access to cookies based on origin, scripts inherit the origin of the main frame.
For example, if \url{tracker.com/myscript.js} runs on \url{example.com}, any cookies it sets are associated with \url{example.com}, granting it full access to all first-party cookies on that site.

\textit{CookieStore API:}
The CookieStore API is a modern interface that provides structured access to cookies via promises.
Unlike \texttt{document.cookie}, it returns cookies as objects, making them easier to query and manipulate.
Available in secure contexts (HTTPS), it is currently supported in Chrome, Safari, and Firefox Nightly~\cite{cookieStore}.
{\sloppy To retrieve a specific cookie, a script can call \texttt{cookieStore.get("cookieName")}.
To access all available cookies, the \texttt{cookieStore.getAll()} function can be used, which behaves similarly to \texttt{document.cookie} but returns a structured array of cookie objects.\par}

\textbf{Cookie categorization.} Browsers traditionally distinguish between first-party and third-party cookies based on the domain attribute. 
However, this model misses a critical category: \textit{ghost-written cookies}—cookies created by third-party scripts executing in the main frame. 
The cookies set by third-party scripts executing in the main frame are treated as first-party cookies. 
As prior work has shown~\cite{sanchez2021journey}, this leads to subtle security and privacy risks. 
In this paper, we focus on all first-party cookies created in the main frame, encompassing both genuine first-party and ghost-written first-party cookies.
\section{Threat Model}
\label{section:threat_model}
Our threat model focuses on third-party scripts embedded in the main frame of a website. 
We consider the security and privacy risks posed by such scripts, which execute with the full privileges of the embedding origin.

\begin{figure}[!thb]
    \centering
    \includegraphics[width=0.65\linewidth]{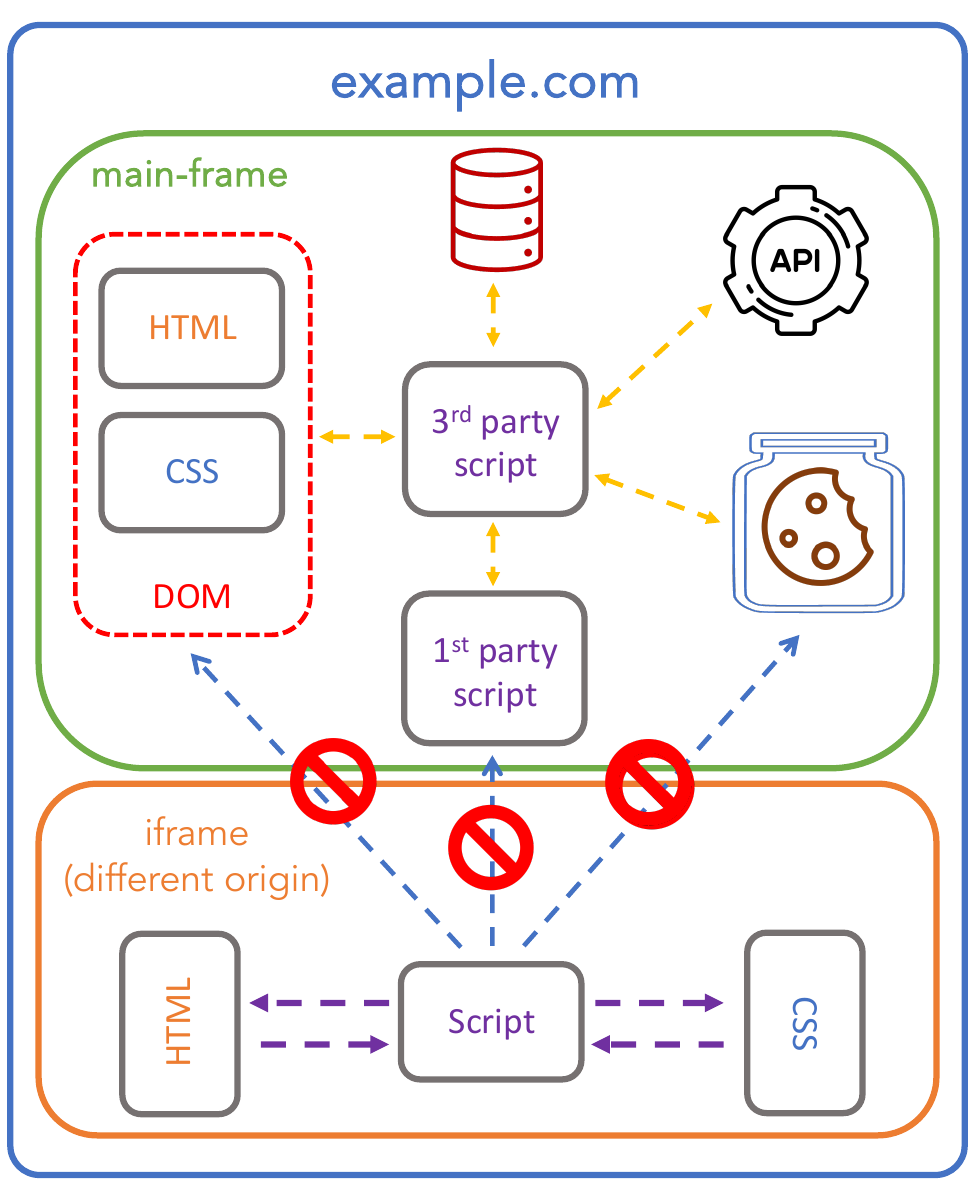}
    \caption{Access of a script included in the main frame of a website to shared resources.}
    \label{fig:threat_model}
\end{figure}

As discussed in Section~\ref{subsec:security_mechanisms}, modern browsers provide baseline defenses against untrusted content. 
CSP mitigates injection by constraining which scripts and other resources may be loaded or executed  but once loaded, CSP does not govern what resources a script can access within a site.
While SOP serves as the browser’s primary defense mechanism for isolating content across origins. 
Under SOP, scripts running in an <iframe> sourced from a different origin than the main frame are prohibited from accessing the main frame’s resources such as DOM and  cookies. 
As illustrated in Figure~\ref{fig:threat_model}, this boundary protects the main frame from untrusted cross-origin iframes, but it does not apply to scripts that execute within the main frame itself. 

As a result, third-party scripts embedded in the main frame inherit all the privileges of the first-party origin, giving them access to the DOM, cookie jar, other included scripts, and additional first-party resources.
Including third-party scripts in the main frame is common in practice, either intentionally (e.g., for analytics or advertising) \cite{metapixel,tiktokpixel} or unintentionally \cite{ruohonen2018integrity} via transitive inclusion chains \cite{ikram2019chain,ikram2020measuring}.
With the phase-out of third-party cookies, trackers have increasingly turned to \emph{first-party cookies}, allowing embedded third-party scripts to use the first-party cookie jar to store and exfiltrate user and device identifiers.
To gain this level of access, trackers' scripts must necessarily execute in the main frame, ensuring it inherits the full privileges of the embedding origin.

% who is adversary
%%
The adversary in our threat model is a third-party script embedded in the main frame, either directly by the developer or indirectly through a transitive inclusion chain.
We explicitly exclude scripts isolated in iframes or sandboxed contexts, since SOP already constrains them from accessing first-party resources.
From the set of main-frame resources, our analysis focuses specifically on cookies accessible via JavaScript; namely, non-HttpOnly HTTP cookies and those set directly through JavaScript. 
We refer to the interactions of this third-party script with the main frame's resources as \emph{cross-domain} interactions.
As we stated in Section~\ref{subsec:security_mechanisms}, we use cross-domain to denote interactions between scripts from different eTLD+1 domains executing in the same main-frame origin, and reserve cross-origin for the stricter SOP definition.
Within this model, an adversarial script may:
(1) \textbf{Read cookies} set by other cross-domain scripts, exposing identifiers such as session tokens, location data, or browsing history. If session cookies are exposed, the adversary can potentially hijack authenticated sessions and impersonate the user~\cite{aliyeva2021oversharing,ghasemisharif2018single}.
(2) \textbf{Overwrite or delete cookies} set by other cross-domain scripts, altering application logic, user state, or enabling persistence cookie-based tracking;
(3) \textbf{exfiltrate cookie contents} by embedding identifiers or values in outbound requests to third-party domains. 
Together, these capabilities demonstrate how cross-domain inclusions in the main frame undermine cookie integrity and pose privacy risks.

Some third-party scripts, such as analytics libraries or personalization services, are intentionally included by developers because functionality such as fine-grained user interaction tracking cannot be achieved from within an iframe. 
However, the fact that certain inclusions are benign does not mean that all included scripts should be implicitly trusted with first-party privileges.
This uniform level of trust is risky even when individual scripts appear benign: an analytics library may inadvertently expose authentication cookies, and a tag manager can transitively load additional malicious scripts or tracking scripts that repurpose first-party identifiers for cross-site tracking. 
By treating all third-party inclusions as fully trusted, current browser security models overlook these risks.

\section{Measurement Methodology} 
To quantify the prevalence of cross-domain cookie interactions in the wild, we design and deploy a custom Chrome extension that performs fine-grained instrumentation of both script- and HTTP-level cookie interactions, as well as outbound network requests. 
Our methodology consists of three key components: (1) a browser extension that performs dynamic instrumentation, (2) a data collection pipeline that crawls popular websites and simulates user interaction, and (3) an analysis framework that identifies cross-domain cookie manipulations and exfiltrations, and the responsible third-party scripts.

\subsection{Extension Instrumentation}\label{web_crawler}
Our Chrome extension enables visibility into both synchronous and asynchronous cookie APIs, server-set cookies, and network requests—all within the main frame of visited pages.

\textbf{Intercepting \texttt{document.cookie} API}
{\sloppy We override the native \texttt{document.cookie} property using \texttt{Object.defineProperty}, wrapping its \textit{getter} and \textit{setter} to intercept all read and write operations. 
For each access, we log the cookie string, the calling script’s URL (derived from the stack trace), and the base domain of the visited site. 
These logs are sent to the extension’s background service via postMessage. \par

\textbf{Intercepting \texttt{CookieStore} API.} 
We override the \textit{get}, \textit{getAll}, \textit{set}, and \textit{delete} methods of the \texttt{CookieStore} API to monitor asynchronous cookie operations. 
Each access is logged with the cookie name, value, method, and initiating script URL.

\textbf{Capturing HTTP \texttt{Set-Cookie} Headers} 
To capture server-set cookies, we use Chrome’s \texttt{webRequest.onHeadersReceived} API to inspect HTTP response headers. 
We extract non-HttpOnly \textit{Set-Cookie} values and log them along with the response URL and the initiator of the request. 
By comparing the initiator and response domains, we determine whether the cookie was set in a first- or third-party context.

\textbf{Network Request Attribution}
To attribute outgoing data transmissions to third-party resources, we attach to each tab using the Chrome Debugger Protocol and listen for \texttt{Network.requestWillBeSent} events. 
When a request is initiated, we analyze the stack trace to identify the exact script responsible. 
This allows us to connect network activity (e.g., exfiltration) to prior cookie accesses, enabling end-to-end attribution.

\subsection{Data Collection}
We crawl the landing pages of the top 20,000 websites from the Universal Tranco list \cite{LePochat2019}. 
All crawls were conducted from a university network located in California, USA.
Page visits are automated using Selenium, which launches a Chrome instance preloaded with our instrumentation extension.

As the page loads, our extension records cookie reads, writes, deletions, Set-Cookie headers, and network requests.
To simulate real user interactions on a website, we perform light interaction by automatically scrolling through the page and clicking on a randomly selected link up to three times, pausing two seconds between each step.
We retain only those sites for which both cookie access logs and network request data are successfully collected.
Applying this criterion, we obtained complete data for 14,917 websites used in our analysis.

Our analysis focuses exclusively on cross-domain interactions that occur in the \emph{main frame}. 
According to the Same-Origin Policy (SOP), scripts executing inside iframes are restricted from accessing the main frame’s DOM or cookie jar unless they share the same origin. 
As such, we restrict our analysis to scripts running in the main frame, where these protections do not apply.

By combining cookie access monitoring with network attribution, our methodology enables a detailed analysis of how third-party scripts manipulate or exfiltrate cookies in real-world websites.
%
% We employed the Universal Tranco list, which is a global top-ranking list, for this purpose. 
%
% It is essential to mention that we conducted prior experiments using alternative methods, such as crawling the top 1,000 websites from 10 different countries and gathering 10,000 Tranco sites filtered to be US-based. 
%
% However, after conducting a comprehensive analysis, we observed no significant differences among these three lists, which ultimately led us to proceed with the Universal list.
%
% Furthermore, we exclusively examined the landing page.
% %
% The homepage stands out as a critical element of every website since it frequently serves as users' initial interaction point.
% %
% When privacy-related information or tracking practices are prominently featured on the homepage, they may extend to subpages as well.
% %
% However, it is important to clarify that our primary objective is not to identify trackers on every subpage. Instead, our aim is to illustrate the privacy consequences associated with the inclusion or injection of third-party scripts into the main frame of websites.%

% Out of the 10,000 websites in our study, PageGraph successfully generates a graph for 67\% of them.
% %
% It is worth noting that while PageGraph's crawler demonstrates a robust success rate of nearly 90\% for the top 1,000 pages, this rate gradually decreases for websites further down the list. 
% %
% This variation in success rates is a limitation of PageGraph.

% \subsection{Command Line Analysis Tool}
\label{section:comman_line_tool}
% To measure the prevalence of cross-domain cookie manipulation and responsible scripts for such actions, we implement a command-based tool in Rust and built on top of PageGraph.
% %
% More precisely, it takes PageGraph generated graphs in the graphml format as input, analyzes page behavior, and generates an output in JSON format.
% %
% Detailed information about the available commands and their usage syntax is provided in Appendix \ref{sec:appendix_pagegraph+}.
% %
% For this paper, we only use \textit{Storage Access} and \textit{Chain of Scripts} queries.

% To study cross-domain cookie access we use our PageGraph+ framework to crawl the main frame of the top 10,000 Tranco-ranked websites. 
% %
% PageGraph+ allows us to intercept how scripts interact with the cookie jar and trace how these scripts are included in the main frame.
\subsection{Identifying Advertising/Tracking Scripts}
\label{section:identifying_tracking}
Throughout this paper, one of our primary objectives is to identify tracking and advertising scripts that are included in the main frame.
To this end, we leverage the adblockparser \cite{adblockparser} tool that is designed for matching URLs against filter rules. 
We combine nine popular crowd-sourced filter lists \cite{EasyList,EasyPrivacy,fanboysocial,fanboyannoyances,Anti_Adblock_Killer.,peterlowes,warning,squid,Blockzilla}, including EasyList and EasyPrivacy, that are specifically curated to target URLs associated with advertising and tracking.
Note that we classify each occurrence of a third-party script URL within a particular website context as advertising/tracking or not.}

\subsection{Analysis Framework}
We analyze the collected logs to detect and attribute cross-domain cookie access, manipulation, and exfiltration by third-party scripts.
The results of this analysis are presented in Section~\ref{sec:analysis}.
Before presenting our findings, we describe the methodology used to detect cross-domain access and exfiltration events.

\textbf{Cross-domain Access Detection.} We label a cookie access as  \textit{cross-domain} if the domain of the accessing script differs from the domain of the script that originally set the cookie.
To identify such cases, we perform the following steps:
\begin{enumerate}
    \item Identify all cookies set via script, along with the domain of the responsible script.
    \item Track subsequent reads or modifications of these cookies by other scripts.
    \item Determine whether those accesses resulted in manipulation (i.e., overwriting or deletion) or exfiltration.
    \item Annotate the inclusion path of each accessing script—whether it was directly embedded by the first-party site or loaded indirectly via third-party chains.
\end{enumerate}

\textbf{Detecting Exfiltration.}
To detect exfiltration of cookie identifiers, we analyze the actual outbound network requests initiated by third-party scripts executing in the main frame, rather than relying on static URL strings. 

Our identifier detection pipeline begins by extracting substrings from cookie values using the format \texttt{(name1 =)value1|...|(nameN =)valueN}, where \texttt{|} represents a delimiter (non-alphanumeric characters). 
We split each value on such delimiters and keep strings that are at least eight characters long, which we treat as candidate identifiers.
For each candidate identifier, we compute three encoded forms: (1) Base64 encoding, (2) MD5 hash, and (3) SHA1 hash. 
We apply a similar algorithm to extract potential identifiers from the query strings of all outbound URLs initiated by third-party scripts in the main frame. 
Exfiltration is confirmed when an identifier derived from a cookie value appears in a request to a different domain.

% \subsection{Identifying Advertising \& Tracking Resources}
% \label{section:identifying_tracking}
% Throughout this paper, one of our primary objectives is to identify tracking and advertising scripts that are included in the main frame.
% %
% To this end, we leverage the adblockparser \cite{adblockparser} tool that is designed for matching URLs against filter rules. 
% %
% We combine nine popular crowd-sourced filter lists \cite{EasyList,EasyPrivacy,fanboysocial,fanboyannoyances,Anti_Adblock_Killer.,peterlowes,warning,squid,Blockzilla}, including EasyList and EasyPrivacy, that are specifically curated to target URLs associated with advertising and tracking.
% %
% Note that we classify each occurrence of a third-party script URL within a particular website context as advertising \& tracking or not.

% \xxx[Umar]{This is the first time, I am reading about advertising/tracking resources? Why do they need to be identified?}
% \input{sections/5_ Measurement_Study_Findings}
\section{Cross-domain First-Party Cookie Access by Third-Party Scripts}
\label{sec:analysis}
In this section, we present the results of our measurement study on cross-domain cookie manipulation and exfiltration, by third-party scripts embedded in the main frame.
We also investigate the predominant scripts responsible for cross-domain actions.
Before that, we report general statistics about script-based cookie access across our dataset to provide context for the cross-domain behaviors analyzed in the following subsections.
\subsection{Prevalence of Third-Party Scripts} Third-party scripts are deeply embedded in the web ecosystem. 
We find that 93.3\% of the 14,917 successfully crawled websites in our dataset include at least one third-party script in their main frame, with an average of 19 distinct third-party scripts per site. 
Notably, 70\% of these scripts are affiliated with advertising or tracking services.

This level of integration has significant implications: on average, third-party scripts set 15 cookies per site, compared to just four set by first-party scripts.
%
% Because browsers grant all main-frame scripts equal privileges, third-party code can access and modify any cookies associated with the first-party origin.

%
\subsection{Usage of Cookie APIs in the Wild}
We measure the usage of both available approaches to set and get script cookies: 1) document.cookie, and 2) cookieStore.

\textbf{document.cookie:} We observe that \texttt{document.cookie} is invoked on 96.3\% of sites. 
Across our dataset, we identified 81,918 unique cookie pairs, each defined as a tuple of (\textit{cookie\_name}, \textit{domain of the script that set the cookie})\footnote{For example, a cookie named \_ga set by \url{google-analytics.com/analytics.js} is represented as (\_ga, google-analytics.com) and is treated as a distinct cookie throughout our analysis.}. These cookies were set by 92,235 scripts originating from 11,035 distinct domains.

\textbf{cookieStore API:} In contrast, the newer \texttt{cookieStore} API \cite{cookieStore} is used far less frequently, appearing on only 2.8\% of sites. 
We identify 411 unique cookie pairs created via \texttt{cookieStore.set()} by 428 scripts across 361 domains.
Despite this diversity, only 13 unique cookie names were used.
Nearly 90\% of these are limited to just two cookie names: \textit{\_awl} and \textit{keep\_alive}.
All instances of keep\_alive are attributed to the Shopify performance SDK (shopify-perf-kit-1.6.X.min.js) hosted on \url{cdn.shopifycloud.com}, suggesting consistent deployment across Shopify-powered sites~\cite{Shopifykeepalive}. 
Similarly, \_awl cookies follow a structured format encoding a counter, timestamp, and session ID (e.g., \_awl=\textit{count.timestamp.session\_id}). 
Although undocumented, manual inspection of setting scripts reveals shared logic: the use of a JavaScript object \texttt{u} with fields like \texttt{u.u = "admiral"}, and references to endpoints on \url{getadmiral.com}, indicating the SDK is part of Admiral’s tracking infrastructure.

%
% \textbf{Security and privacy implications of main-Frame execution}
% The equal privilege model for main-frame scripts introduces a fragile and overexposed trust boundary. 
% %
% Third-party scripts, despite their distinct origin, can freely access and manipulate cookies set by others. 
% %
% This model significantly expands the attack surface for unauthorized tracking and manipulation.

% %
% This threat is not merely theoretical. 
% %
% For instance, Chen et al.~\cite{chen2021cookieswap} report a case where a script from Adobe’s tag management service (\url{assets.adobedtm.com}) accessed a \_ga cookie originally set by Google Analytics and exfiltrated it to Adobe’s tracking domain (\url{demdex.net}) on \url{Uplus.co.kr}. 
% %
% Such access enables cross-site user tracking by correlating identifiers across unrelated domains.

\subsection{Prevalence of Cross-domain Cookie Interactions}
While prior work has extensively examined cookie exfiltration, less attention has been paid to other forms of manipulation such as overwriting and deletion. 
We now systematically quantify three categories of cross-domain actions: \textbf{exfiltration}, \textbf{overwriting}, and \textbf{deletion}. 
We begin by quantifying the prevalence of \emph{cross-domain} cookie interactions for cookies set via the \texttt{document.cookie} property and \texttt{cookieStore} API.

As shown in Table~\ref{tab:unauthorized-cookie-actions}, cross-domain interactions to cookies set through \texttt{document.cookie} is both widespread and varied in nature. 
In contrast, such access to cookies set via the \texttt{cookieStore} API is significantly less common.
\begin{table}[h]
\centering
\resizebox{\columnwidth}{!}{%
\begin{tabular}{@{}llrr@{}}
\toprule
\textbf{cookie Type}                     & \textbf{Action} & \textbf{\% of Websites} & \textbf{\% of Cookies (No.)} \\ \midrule
\multirow{3}{*}{\textbf{document.cookie}} & exfiltration    & 55.7                    & 5.9 (4,825)                  \\
 & overwriting & 31.5 & 2.7 (2,212) \\
 & deleting    & 6.3  & 1.8 (1,475) \\ \midrule
\multirow{3}{*}{\textbf{cookieStore}}    & exfiltration    & 0.7                     & 16.3 (62)                    \\
 & overwriting & 0    & 0           \\
 & deleting   & 0    & 0           \\ \bottomrule
\end{tabular}%
}
\caption{Prevalence of cross-domain cookie actions across websites and affected cookies.}
\label{tab:unauthorized-cookie-actions}
\end{table}

\textbf{Cross-domain exfiltration.}
Cross-domain cookie exfiltration is the most prevalent forms of cookie misuse.
We find that about 55.7\% of websites include at least one script that exfiltrates a cookie it did not set originally.
These behaviors impact 5.9\% (4,825) of the 
~82,000 unique cookie pairs in our dataset, highlighting the scale at which confidentiality is compromised in the main frame.
For cookies set via the \texttt{cookieStore} API, cross-domain exfiltration is significantly less common. 
Only 0.7\% of websites include a script that exfiltrates a \texttt{cookieStore}-set cookie it did not set originally. 
These scripts affect 16.3\% (62) of the 481 unique \texttt{cookieStore} cookie pairs observed in our dataset.

\textbf{Cross-domain manipulation.}
Cross-domain manipulation actions (including cookie overwriting or deleting) are less common but still pose meaningful integrity risks. 
We find that approximately 32\% of websites include at least one script that overwrites a cookie it did not originally set, while 6.3\% include a script that deletes a cookie it did not create. 
Overall, 2.72\% of all unique cookie pairs are overwritten by non-owning scripts, and 1.8\% are explicitly deleted.
For cookies set via the \texttt{cookieStore} API, we did not observe any cross-domain overwriting or deletion.
% In many cases, these actions target cookies with high entropy or non-guessable names—suggesting that the scripts have full access to the cookie jar and could deliberately modify stateful identifiers.

Together, these findings underscore the fact that cookie access in the main frame lacks effective isolation and is routinely violated. 
In the following subsections, we examine notable instances of these cross-domain actions and attribute them to the responsible script origins.
Due to its negligible usage, the \textit{cookieStore} API is excluded from further analysis.

%% ########################
\subsection{Cross-domain Cookie Exfiltration}
While authorized exfiltration by scripts from the same origin is a widespread and expected practice—for example, for analytics, session management, or personalization—cross-domain exfiltration lacks visibility and control, raising serious privacy concerns.

There are currently no built-in browser controls to restrict what data a script exfiltrates once it has access to the cookie jar. 
As a result, cookies containing user identifiers, authentication tokens, and user browsing data are routinely extracted and sent to unrelated third-party domains~\cite{dimova2021cname, huo2022all, farooqi2020canarytrap}.

%%%%%%%%%%%%%%%%%%%
\textbf{Most Commonly Exfiltrated Cookies.} Table~\ref{tab:victim_cookies_exfiltration} lists the top 20 unique cookies most frequently exfiltrated by cross-domain scripts.
To highlight real-world privacy risks and consolidate domains belonging to the same entity, we map the URLs of cookie-exfiltrating scripts and their destination endpoints to their respective entities using DuckDuckGo's Tracker Radar dataset~\cite{DDGTrackerRadar}.
We find that the \_ga cookie—created by scripts embedded from \texttt{googletagmanager.com}—is the most frequently exfiltrated cookie. 
Scripts from 1,191 distinct entities (excluding Google) accessed and transmitted this cookie to 708 unique recipient entities. 
Other frequently targeted cookies include \_gid, \_gcl\_au, and OptanonConsent, often associated with major analytics platforms. 
Microsoft, Yandex, and Pinterest appear as top exfiltrators, while HubSpot, Amazon, and Microsoft are the most frequent exfiltration destinations.

One notable exception in Table~2 is the \texttt{us\_privacy} cookie, which encodes the IAB U.S. Privacy (CCPA) signal. 
Unlike tracking identifiers, this cookie is \emph{intended} to be read and exfiltrated by third parties so that downstream ad tech can honor a user’s “Do Not Sell” preference.\footnote{In practice, consent strings are commonly stored in a cookie (often named \texttt{us\_privacy} or \texttt{usprivacy}) and read by third parties implementing the IAB signal.} We therefore flag \texttt{us\_privacy} as a \emph{consent signal} rather than a tracking identifier. Although prior work finds that their implementation is not always compliant \cite{petsIAB2024,iabCCPAtech}.

\textbf{Case study: Targeted cookie parsing and exfiltration.}
We analyze a concrete example of targeted exfiltration on \url{optimonk.com}, where a third-party script loaded from \texttt{licdn.com/li.lms-analytics/insight.min.js} (LinkedIn) accesses and processes the \_ga cookie created by \texttt{googletagmanager.com}. 
The \texttt{\_ga} cookie on \texttt{optimonk.com} holds the value:
\texttt{GA1.1.444332364.1746838827}.

This format is consistent with Google Analytics, where the middle segment (\texttt{444332364}) represents a pseudonymous identifier. 
The script extracts this segment, encodes it in Base64 (\texttt{NDQ0MzMyMzY=}), appends an additional encoded timestamp (\texttt{0LjE3NDY4Mzg4Mjc}), and transmits it to a LinkedIn endpoint via a GET request to:

\begin{mybox}
\small
\raggedright
\url{https://px.ads.linkedin.com/attribution\_trigger?pid=621340\&time=1746838846149\&url=www.optimonk.com*\_gcl\_au*MTg5MTM3M4xNzQ2ODM4ODI3*FPAU*...*\_ga*NDQ0MzMyMzY0LjE3NDY4Mzg4Mjc...}
% \texttt{https://px.ads.linkedin.com/attribution\_trigger?
% \\
% pid=621340\&time=1746838846149\&url=https\%3A\%2F\%2F
% \\
% www.optimonk.com\%2Fklaviyo\%2F\%3F\_gl\%3D1*1e8pfou
% \\
% *{\textbf{\_gcl\_au}}*MTg5MTM3Mjkw...*{\textbf{FPAU}}*MTg5MTM3Mj...*
% \\
% {\textbf{\_ga*NDQ0MzMyMzY0.LjE3NDY4Mzg4Mjc}}*{\textbf{\_fplc}}*emNrU1R...}
\end{mybox}

A closer inspection of the URL reveals the presence of several additional cookie-derived identifiers, including \texttt{\_ga}, \texttt{\_fplc}, \texttt{gcl\_au}, and \texttt{fpau}, suggesting general tracking intent. 
Notably, although over 35 cookies were present during the visit, the script selectively parsed and exfiltrated only a subset—suggesting that cookie exfiltration here is purposeful. 
This undermines the assertion that identifier exfiltration here is merely a byproduct of bulk cookie access.

\textbf{Case Study: Cross-company identifier exfiltration.}
We also identify a case of cross-company identifier exfiltration.
We observe a case of identifier sharing between Facebook and Criteo facilitated through a third-party script on the website \url{goosecreekcandle.com}. 
A script hosted by Osano, Inc. \cite{osano}, a consent management provider, accesses a Facebook cookie (\_fbp) and transmits it to Criteo SA's infrastructure. Specifically, \texttt{osano.com/1vX3GkPazR/.../osano.js} accesses the \texttt{\_fbp} cookie set by \texttt{facebook.net}: \texttt{fb.0.1746746266109.868308499845957651}

The script extracts two core components: (1) \texttt{1746746266109}, a timestamp; and (2) \texttt{868308499845957651}, a Facebook-assigned browser identifier. It then includes both values in a request to:

\begin{mybox}
\small
\raggedright
\url{https://sslwidget.criteo.com/event?...&sc=\%7B\%22fbp\%22:\%22fb.1.1746746266109.868308499845957651\%22...}
\end{mybox}

This identifier transfer occurs within the context of a   partnership between Meta and Criteo \cite{CriteosComplaint,CriteosContractfacebook}.
Given this partnership, we surmise that the data exchange is likely intentional. 
Criteo may receive the \texttt{\_fbp} identifier to help align its own user profiles with Meta’s for improved cross-platform attribution and conversion tracking.
While this identifier sharing is between two partners, its execution still raises important privacy considerations. 
Users are typically unaware that consent interfaces or third-party privacy scripts (like Osano) may enable identifier syncing between advertisers/trackers. 
Moreover, the fact that a consent management tool also facilitates data transmission underscores the blurred lines between privacy-enhancing and tracking technologies.

\begin{table*}[!htb]
\centering
\scriptsize
\begin{adjustbox}{max width=\textwidth}
\begin{tabularx}{\textwidth}{@{}l l r r X X@{}}
\toprule
\textbf{Cookie Name} & 
\textbf{Owner Domain} & 
\makecell{\textbf{\# Exfiltrator}\\\textbf{Entities}} & 
\makecell{\textbf{\# Destination}\\\textbf{Entities}} & 
\makecell{\textbf{Top 3}\\\textbf{Exfiltrator Entities}} & 
\makecell{\textbf{Top 3}\\\textbf{Destination Entities}} \\
\midrule
\_ga & googletagmanager.com & 1191 & 664 & Microsoft, Yandex, Pinterest & HubSpot, Microsoft, Amazon \\

\_gid & google-analytics.com & 718 & 576 & Microsoft, Pinterest, Yandex & HubSpot, Microsoft, Yandex \\

\_ga & google-analytics.com & 482 & 456 & Microsoft, Yandex, Pinterest & HubSpot, Microsoft, Yandex \\

\_gcl\_au & googletagmanager.com & 586 & 416 & Microsoft, Pinterest, Adobe & HubSpot, Microsoft, Yandex \\

\_gcl\_au & googletagmanager.com & 367 & 285 & Microsoft, Adobe, HubSpot & HubSpot, Microsoft, Amazon \\

i & openx.net & 157 & 226 & Google, Mediavine, AdThrive & Amazon, LiveIntent, Yandex \\

pd & openx.net & 138 & 217 & Google, Mediavine, AdThrive & Amazon, LiveIntent, Yandex \\

SPugT & pubmatic.com & 87 & 181 & Google, Taboola, AdThrive & Microsoft, Amazon, X \\

PugT & pubmatic.com & 63 & 167 & Google, script.ac, Amazon & Amazon, Yandex, LiveIntent \\

\_\_utma & google-analytics.com & 54 & 159 & Yandex, Microsoft, Adobe & HubSpot, Microsoft, Yandex \\

\_fbp & facebook.net & 73 & 154 & Criteo, Google, Salesforce.com & Criteo, whitesaas.com, c99.ai \\

\_\_utmb & google-analytics.com & 44 & 152 & Yandex, Microsoft, envybox.io & HubSpot, Microsoft, Yandex \\

\_\_utmz & google-analytics.com & 44 & 152 & Yandex, Microsoft, envybox.io & HubSpot, Microsoft, Yandex \\

\_mkto\_trk & marketo.net & 44 & 127 & c99.ai, ensembleiq.com, SSL & Adobe, c99.ai, insent.ai \\

\_ym\_d & yandex.ru & 86 & 126 & Google, mango-office.ru, envybox.io & Yandex, Criteo, whitesaas.com \\

lotame\_domain\_check & crwdcntrl.net & 31 & 114 & Amazon, hadronid.net, Google & Amazon, Criteo, LiveIntent \\

us\_privacy & ketchjs.com & 17 & 112 & Google, tradehouse.media, ex.co & 33Across, Anview, LiveIntent \\

\_yjsu\_yjad & yimg.jp & 30 & 109 & Google, Taboola, Microsoft & Microsoft, Criteo, Adobe \\

gaconnector\_GA\_Client\_ID & gaconnector.com & 4 & 106 & Google, Microsoft, HubSpot & Microsoft, HubSpot, Airbnb \\

gaconnector\_GA\_Session\_ID & gaconnector.com & 4 & 106 & Google, Microsoft, HubSpot & Microsoft, HubSpot, Airbnb \\

sc\_is\_visitor\_unique & statcounter.com & 18 & 103 & Google, Mediavine, Intergi & LiveIntent, Magnite, ShareThis \\
\bottomrule
\end{tabularx}
\end{adjustbox}
\caption{Top 20 cookie names and their creator domains that have been exfiltrated by cross-domain scripts across 20,000 websites. The table also reports the number of cross-domain exfiltrator entities responsible for the exfiltration and the number of destination entities that received these cookies. Entries are sorted by the number of destination entities.}
\label{tab:victim_cookies_exfiltration}
\end{table*}

\begin{figure}[!htbp]
    \centering
    \includegraphics[width=0.8\columnwidth, trim=0 5 0 8, clip]{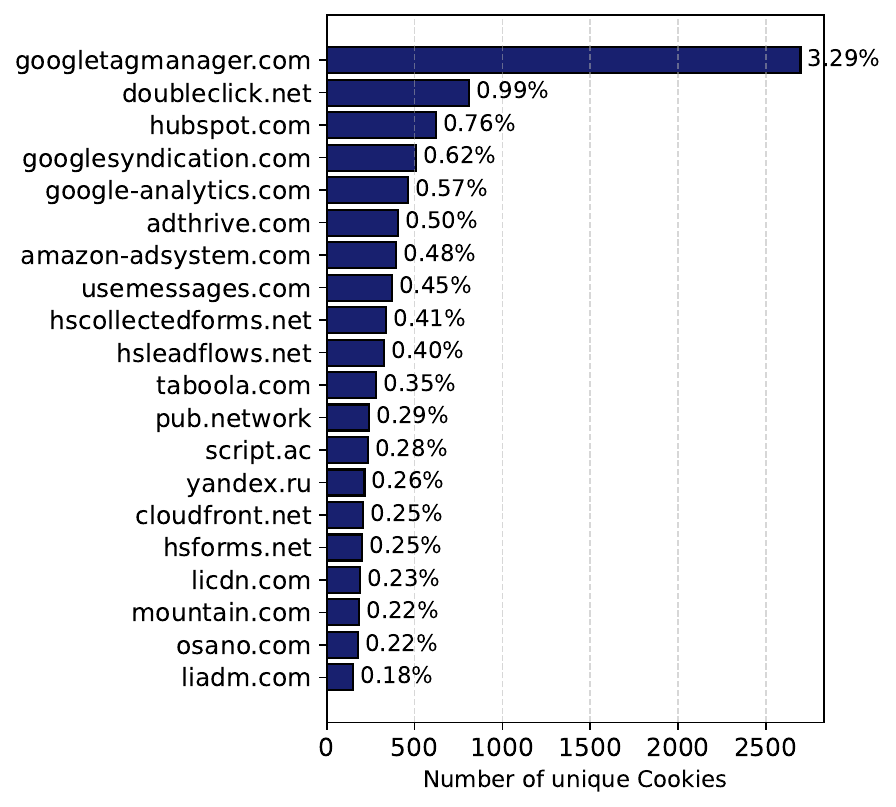}
    \caption{Top 20 script-hosting domains involved in cross-domain cookie exfiltration, ranked by the number of unique cookies exfiltrated to third-party destinations.}
    \label{fig:top_exfiltrator}
\end{figure}

%%%%%%%%%%
\textbf{Top exfiltrator Domains}
Figure \ref{fig:top_exfiltrator} presents the top 20 script domains engaged in cross-domain cookie exfiltration by the number of unique cookies they leaked.

\url{google-analytics.com} is the top exfiltrator, accounting for exfiltrating 3.3\% of the 82,000 cookies in our dataset. 
This domain is primarily associated with analytics services aiming to analyze user behavior by collecting and exfiltrating information.
Several advertising exchanges platforms such as \url{doubleclick.net}, \url{amazon-adsystem.com}, and \url{pubmatic.com}, are also among the top exfiltrators.
Their presence is largely explained by how Real-Time Bidding (RTB) systems operate\cite{olejnik2016bid}. 
During RTB auctions, bid requests are sent to multiple potential bidders and may include user identifiers and cookies—regardless of the script that originally created them. 
With 14.3\% of top Tranco sites supporting RTB~\cite{pachilakis2019no}, widespread cookie leakage to ad networks is expected which raises concerns about cross-domain access and misuse of user data by cross-domain scripts.

\subsection{Cross-domain Cookie Overwriting and Deletion} 
In this section, we focus on cross-domain cookie manipulation including overwriting and deleting.
Invoking the \texttt{document.cookie} API returns the entire cookie jar, regardless of whether the caller script requires all cookies or not.
However, to overwrite or delete a specific cookie value, a script needs to know the corresponding cookie name (key).
As shown in Table~\ref{tab:unauthorized-cookie-actions}, 2,121 and 1,475 of unique cookies are overwritten and deleted by scripts that did not originally set them.
To better understand the nature of cookie overwriting, we analyze the attributes modified during such events. Specifically, we examine changes to the cookie's value, expires, domain, and path fields. Among all overwrite events involving cookies identified as victims of third-party modification, 85.3\% involved a change in the value, 69.4\% updated the expires timestamp, 6.0\% altered the domain, and 1.2\% changed the path. These results suggest that overwriting is primarily used to manipulate the content and lifespan of cookies, potentially to repurpose user identifiers or extend tracking durations beyond the original intent.

\textbf{Most Frequently Modified Cookies.} 
Table~\ref{tab:manipulation-cookie-attacks} in Appendix \ref{appendix:top_manipulator} presents the top 10 unique cookie pairs that are manipulated—either overwritten or deleted—by cross-domain scripts.
The \_fbp cookie, set by \url{facebook.net}, is the most frequently overwritten, targeted by scripts from 132 entities, and also deleted by 18 distinct entities.
Other commonly overwritten cookies include OptanonConsent, \_ga, and cto\_bundle, while frequently deleted ones include tracking-related cookies such as \_uetvid, \_ga, and \_gid.
Scripts from Function Software, Google, and Gatehouse Media dominate cross-domain overwriting actions, whereas cdn-cookieyes.com, cookie-script.com, and Tealium are leading in cross-domain deletion activities.
%
%placeholder for table top manipulators
%

\textbf{Case Study: Intention behind manipulations}
Understanding the intent behind cross-domain cookie manipulations is inherently challenging due to the lack of documentation or explicit disclosures.
Nevertheless, our analysis reveals several recurring patterns and concrete cases that help illuminate the purposes behind these actions.

\textit{Collision.} One possible explanation for this behavior is the use of generic cookie names, such as \textit{cookie\_test} or \textit{user\_id}, which are more likely to be targeted. 
For instance, our analysis identifies at least eight distinct \textit{cookie\_test} cookies—each set by a different script—yet overwritten or deleted by more than 70 unique scripts.
Therefore, these modifications or deletions may result from unintentional name collisions.

\textit{Collusion or Competition.} We observe instances where cookies with non-trivial, hard-to-guess names are overwritten by cross-domain scripts.
For instance, the value of \textit{cto\_bundle} cookie initially created by Criteo is over-written by a script associated with Pubmatic. 
It changes from a string in hash format with a length of 194 characters to an entirely different value in hash format with a length of 258 characters.
This behavior suggests a deliberate overwrite rather than an accidental collision. 
One possible explanation is technical coordination between ad tech companies—such as ID synchronization or shared identifier infrastructure. 
However, such behavior may also signal competitive dynamics, where one entity seeks to override or disrupt the identifiers of another to gain control over user tracking or ad targeting capabilities.

\textit{Privacy Compliance.}
Site owners or tag managers may configure services to delete cookies from certain third parties (e.g., Bing) to enforce compliance with regional privacy laws (e.g., GDPR/CCPA) or to avoid unnecessary data retention \cite{cookieconsent}.

In Table \ref{tab:manipulation-cookie-attacks} we can see \url{cookie-script.com} and \url{cdn-cookieyes.com} among top entities engaged in cross-domain deleting.
Both \url{cookie-script.com} and \url{cookieyes.com} are associated with cookie consent management platforms designed to help websites comply with data privacy regulations like GDPR and CCPA.
Scripts from these services delete tracking cookies such as \_fbp and \_uetvid to enforce user privacy preferences and comply with regulations like GDPR and CCPA. 
If a user declines consent for marketing cookies, these scripts remove pre-existing tracking identifiers to prevent unauthorized data collection.
In some cases, scripts may also delete and overwrite these cookies to reset tracking sessions or reassert control over user identifiers.

\textbf{Top manipulator.} Figure \ref{fig:top_manipulator_domains} in Appendix \ref{appendix:top_manipulator_domains} presents the top 20 script-hosting domains involved in cross-domain cookie overwriting and deletion.
The y-axis lists the top domains, while the x-axis indicates the number of unique first-party cookies manipulated by scripts from each domain.
\url{googletagmanager.com} ranks highest in overwriting activity, modifying 0.5\% of all cookies in our dataset (386 out of 82,000).
On the other hand, \url{prettylittlething.com} leads in cookie deletion, removing 0.31\% of cookies (252 out of 82,000).
Most domains associated with overwriting are tied to advertising and tracking services, whereas those responsible for deletion are primarily linked to consent management platforms.

% \begin{figure}[!htbp]
%     \centering
%     \includegraphics[width=\columnwidth]{plots/attackers/overwriting_top_attacker.pdf}
%     \caption{Top 10 cookie overwriters sorted based on the number of their actions on cookies.}
%     \label{fig:top_overwriterting_attacker}
% \end{figure}

% \begin{figure}[!htbp]
%     \centering
%     \includegraphics[width=\columnwidth]{plots/attackers/deleting_top_attacker.pdf}
%     \caption{Top 10 cookie deleter! sorted based on the number of their actions on cookies.}
%     \label{fig:top_deleting_attacker}
% \end{figure}

\subsection{Where Do These Scripts Come From?}
We classify cookie-accessing scripts by inclusion path:
93.3\% of websites include third-party scripts in the main frame.
Indirect inclusions outnumber direct inclusions by a factor of 2.5×.
33\% of indirect third-party scripts are associated with advertising or tracking.

% \begin{table}[h]
% \centering
% \begin{tabular}{lcc}
% \toprule
% \textbf{Domain} & \textbf{Direct Inclusion} & \textbf{Indirect Inclusion} \\
% \midrule
% googletagmanager.com & 10\%   & 90\%   \\
% doubleclick.net      & 8.6\%  & 91.4\% \\
% \bottomrule
% \end{tabular}
% \caption{Distribution of direct and indirect inclusions for selected domains.}
% \label{tab:inclusion-types}
% \end{table}

Websites often include tools like Google Tag Manager or advertising SDKs directly, which then dynamically inject additional scripts—including those that perform cross-domain cookie operations. 
This inclusion indirection makes it hard for website owners to audit or control access to the cookie jar.

\subsection{Implications}
Our analysis reveals that first-party cookie integrity is fundamentally broken in the presence of third-party scripts in the main frame:
Scripts frequently access, modify, and exfiltrate cookies they did not create.
Such actions occur on nearly half of all websites.
Indirect inclusion chains obscure the origin of cross-domain behavior.
Even potentially sensitive domains (e.g., healthcare) are not immune.
Emerging practices like server-side tracking bypass client-side defenses, including our own \name (§6), by proxying exfiltration through seemingly first-party endpoints \cite{vekaria2025sok,fouad2024devil,amieur2024client}.
These findings call for stronger ownership semantics for browser cookies—where access control is tied to the script that created the cookie, not merely the domain from which it is served.

% April 17
\section{\name: \\ Design \& Implementation} 
\label{label:cookie_guard_design}
Our measurement and analysis in Section~\ref{sec:analysis} reveals that first-party cookies are vulnerable to cross-domain manipulation and exfiltration by third-party scripts embedded in the main frame. 
To mitigate this issue, we introduce \name, a browser extension that enforces isolation of first-party cookies on a per-script-origin basis.

\subsection{Design} \label{sec:cookie_guard_design}
\name enforces access control over first-party cookies by maintaining a metadata store that logs each cookie’s name and the ETLD+1 of the script or server that created it.
This metadata is updated during cookie creation events, both via JavaScript (\texttt{document.cookie} and \texttt{cookieStore.set()}) and HTTP Set-Cookie headers.

When a script attempts to read from document.cookie, \name identifies the script's ETLD+1 and filters the returned cookies, exposing only those that were originally created by scripts from the same domain. 
This policy effectively prevents third-party scripts embedded in the main frame from accessing or interfering with cookies set by other parties.

On some websites, essential functionality like user session management, preferences, or shopping carts are handled by third-party scripts embedded as first party.
Blocking cookies set by these scripts could disrupt website functionality.
To prevent such disruptions, in \name's implementation, we grant full access control to the website owner to get access to all first-party cookies.
Consequently, if a script from the same domain as the visting website invokes \texttt{document.cookie} or \texttt{cookieStore.getAll()}, it retrieves all first-party cookies from the cookie jar, regardless of the domain that set them.

Inline scripts present an attribution challenge, as their origin cannot be reliably determined.
To address this, \name supports two modes of handling inline scripts.
In strict mode, \name adopts a safe-by-default strategy by treating such scripts as untrusted and denying them access to any cookies. 
This conservative stance aligns with privacy defenses that restrict ambiguous behaviors by default to minimize the attack surface and increase the cost of circumvention.
In relaxed mode, by contrast, \name treats inline scripts as first-party scripts. 
We will not use the relaxed mode in our evaluations and it is included only to illustrate alternative design choices.

\begin{figure*}[htb!]
    \centering
    \includegraphics[width=0.9\linewidth]{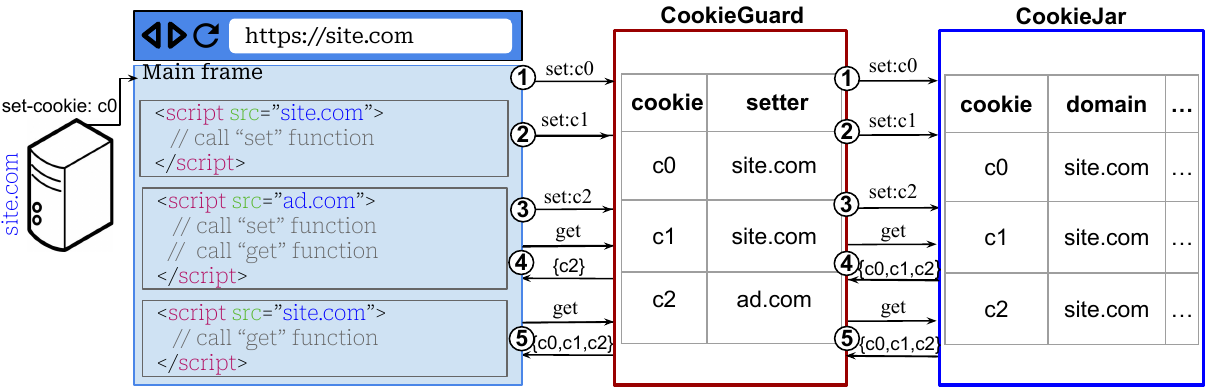}
    \caption{\name overview}
    \label{fig:cookieGuard}
\end{figure*}

Figure \ref{fig:cookieGuard} shows an overview of \name.
\circled{1} A server at \url{site.com} sets cookie ``c0'' via an HTTP \texttt{Set-Cookie} header. Since the response matches the visited domain, the cookie is marked first-party. The domain of the URL in the original cookie jar and creator domain of this cookie in \name's database are both set to \url{site.com}.
\circled{2} A script from \url{site.com} sets cookie ``c1'' using \texttt{document.cookie}. Like ``c0'', it is considered a first-party cookie with \url{site.com} as its creator in both the browser and \name's records.
\circled{3} A third-party script from \url{ad.com}, embedded in the main frame of \url{site.com}, sets cookie ``c2''. While the browser treats it as first-party (domain: \url{site.com}), \name records \url{ad.com} as its creator.
\circled{4} When the script from \url{ad.com} calls \texttt{document.cookie}, the browser returns all first-party cookies (``c0'', ``c1'', and ``c2''). \name filters out ``c0'' and ``c1'' since their creators differ, and returns only ``c2''.
\circled{5} A script from \url{site.com} calls \texttt{document.cookie}. Under the full-access policy (Section~\ref{label:cookie_guard_design}), \name returns all first-party cookies, since the script’s domain matches the visited domain.

In the following section we explain the components of \name, and we detail its implementation.

% In this section, we describe the design and implementation details of NoT.JS. Figure 3 provides an overview of NoT.JS’s pipeline, which
% starts with
% \name has three main features.
% %
% 1. Script Cookie Access Interception: Monitors access to document.cookie to capture get and set operations initiated by web page scripts.

% 2. Monitor set-cookie header in HTTP responses: Monitors all http responses that include a set-cookie header. 

% 3. Bookkeeping: implements meticulous logging for all first-party cookies, encompassing both HTTP and JavaScript cookies. More specifically we keep all first party cookies’s name alongside their setter domain.

\subsection{Implementation of \name}
\begin{figure}[htb!]
    \centering
    \includegraphics[width=0.65\linewidth]{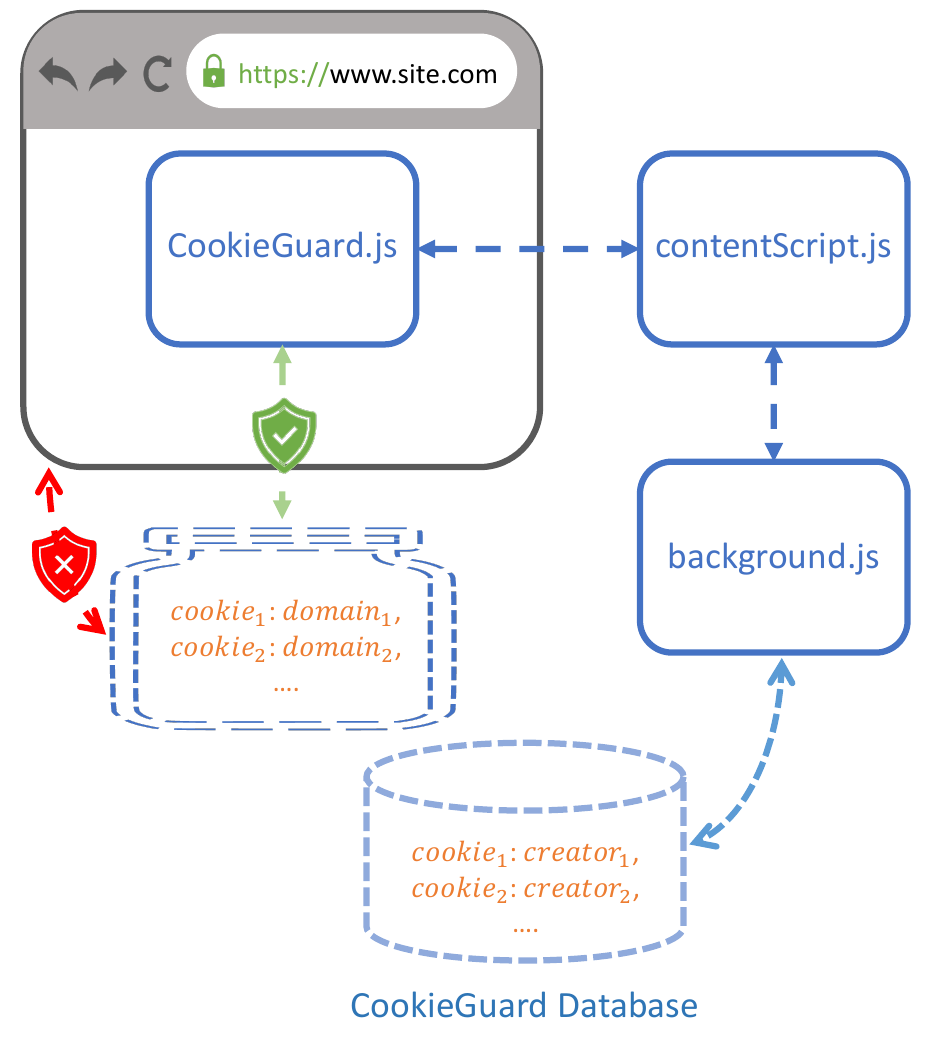}
    \caption{\name's components}
    \label{fig:extension}
\end{figure}

We implement \name as a browser extension, as illustrated in Figure \ref{fig:extension}.
The extension has three main components:

\texttt{contentScript.js}: Injects \texttt{cookieGuard.js} into the main frame and mediates communication between \url{cookieGuard.js} and \texttt{background.js}. 
It relays messages to ensure synchronization between DOM-level script activity and background-level cookie tracking.

\texttt{cookieGuard.js}: Intercepts cookie access by wrapping both the legacy \texttt{document.cookie} and the modern \texttt{cookieStore} API. 
It implements a wrapper around the \texttt{document.cookie} and \texttt{cookieStore}'s  \textit{get()} and \textit{set} functions, enabling the interception of cookie jar interactions. 
For cookie writes, \texttt{cookieGuard.js} captures the cookie name and the domain of the calling script, inferred by analyzing the JavaScript stack trace to locate the last external script URL. 
This metadata is sent to \textit{background.js} to update the extension’s internal dataset.
For reads, \texttt{cookieGuard.js} queries background.js for the latest dataset and filters the cookies based on the ETLD+1 of the caller. If the script’s domain matches the top-level site, all first-party cookies are returned (preserving default browser behavior). Otherwise, only cookies set by the same third-party domain are exposed.

\texttt{background.js}: Monitors HTTP responses that contain Set-Cookie headers, logging details of non-HTTPOnly and first-party cookies by name and setter domain in the extension's storage. 
It handles all updates to the dataset initiated by 'set' requests from both HTTP headers and \texttt{cookieGuard.js}. 
Furthermore, when the 'get' function is invoked in \texttt{cookieGuard.js}, \texttt{background.js} receives a message through \texttt{contentScript.js} to provide a current copy of the dataset for accurate cookie filtering and response.     

%

% \subsubsection{Implementation}
% Cookie Interception: The background script (background.js) is tasked with intercepting all network requests that include a Set-Cookie header without the HttpOnly attribute. It constructs a cookie object containing the setter's domain, the cookie's domain (extracted from the header), and the entire cookie object. This information is then used to update the extension's cookie database.

% %
% Cookie Setting via document.cookie: Implemented in cookieProtect.js, this functionality records all calls to the setter function of document.cookie, capturing the domain of the setter, the name of the cookie, and its value. This data is sent to background.js to keep the cookie database current, ensuring our system's integrity and effectiveness in protecting user privacy.

% Domain attribute matching: Cookies set by a script embedded within the mainframe of the website, characterized by a domain attribute that aligns with the website's domain, are logged and protected. Our tool ensures that only the scripts from the originating domain have access to these cookies, reinforcing domain-level isolation and privacy.
\section{\name Evaluation}% April 19
% twitter breakage example => disconnect.me
% www.kirkusreviews.com => white list
We assess the efficacy of \name, our cookie protection tool, through a multi-faceted evaluation framework, focusing on accuracy, website breakage, and runtime performance.

\subsection{Access Control}
\begin{figure}[!htb]
    \centering
    \includegraphics[width=0.5\linewidth, trim=0 5 0 5, clip]{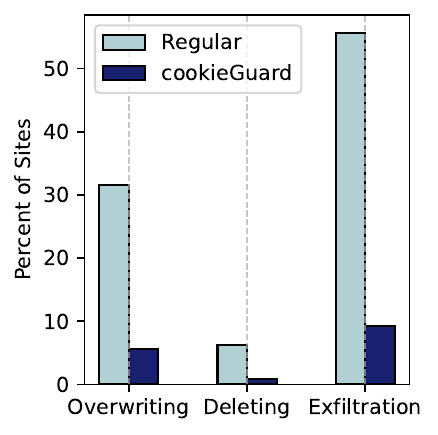}
    \caption{Evaluation of cross-domain cookie interactions: Comparing regular browser behavior with and without the \name extension.}
    \label{fig:evaluation}
\end{figure}

We assess the effectiveness of \name in preventing cross-domain actions such as the retrieval, overwriting, deletion, and exfiltration of first-party cookies. 
Figure \ref{fig:evaluation} illustrates the comparison between the percentage of sites engaging in these actions on first-party cookies with the \name extension enabled versus a regular browser without the extension.
Notably, \name significantly reduces cross-domain actions on first-party cookies: overwriting by 82.2\%, deletion by 86.2\%, and exfiltration by 83.2\%.
As discussed in Section~\ref{sec:cookie_guard_design}, to minimize website breakage, \name grants website owners full control over access to first-party cookies by their own scripts.
As a result, cross-domain actions on first-party cookies may still occur under \name, and the percentage of affected sites is not zero.

\subsection{Website Breakage}
We perform a manual assessment of \name to determine its impact on website functionality when the \name extension is enabled.
A random sample of 100 websites from the Tranco top 10k list is selected and tested in two separate settings: with and without the \name extension activated.
One author examines all 100 websites, while two additional evaluators are recruited to independently assess website breakage.
Each of these evaluators is responsible for assessing a unique set of 50 websites from our chosen sample.
At least two distinct evaluators review each website to ensure a comprehensive evaluation.

We classify breakage into four categories: navigation (moving between pages), SSO (initiating and maintaining login state), appearance (visual consistency), and other functionality (such as chats, search, and shopping cart). 
Breakage is labeled as either major or minor for each category: Minor breakage occurs when it is difficult but not impossible to use the functionality. 
Major breakage occurs when it is impossible to use the functionality on a web page.

The results are shown in Table \ref{tab:web-breakage}.
While \name has no impact on navigation and appearance (0\% breakage), we observe 1--11\% minor and major breakage on SSO and other functionality.
More specifically, on 11\% of the websites, we are not able to sign in using SSO, because the sign-in process is managed by third parties. 
When the user logs into SSO provider's system such as Google account, the system authenticates the user and upon successful authentication, the SSO provider sets a session cookie in the user's browser. 
This cookie is then used to remember the user's authentication status. 
The SSO that leads to web page breakage stems from the dependency on third-party cookies for session management and login process. 
On some websites, multiple third-party scripts are in charge of session management. 
For example, on \url{zoom.us} two scripts, i.e. \url{microsoft.com} and \url{live.com} are in charge of SSO.
Since \name eliminates all cross-domain manipulations, in such cases breakage happens.
An example of minor breakage occurs on \url{cnn.com}, where a user can sign in, but refreshing the page logs them out.
This behavior results from a limitation in our browser extension implementation, which interferes with how certain session cookies are handled across page reloads.
In terms of functionality, on 3\% of the websites, there is at least one advertisement served by a third-party script, which is not shown due to our extension; and in another 3\% of the websites, \name leads to major breakage in functionality.
For example, on \url{facebook.com}, Facebook Messenger, an instant messenger service owned by Facebook, is managed by cookies served from \url{fbcdn.net}. 
Although \url{fbcdn.net} is a CDN belonging to Facebook, since their domains are different, it is considered as third-party on \url{facebook.com}.
Therefore, its access to the cookie jar is limited by \name which leads to not working properly.
To address compatibility issues on \url{facebook.com} and other websites with similar behaviors, we implement a whitelist feature in \name that groups all domains belonging to the same entity, using information from DuckDuckGo’s entities list~\cite{DDGTrackerRadar}.
This refinement reduces website breakage to 3\%.

\begin{table}[!tbh]
% \small
\centering
\small
\begin{tabular}{@{}lcccc@{}}
\toprule
 & \textbf{Navigation} & \textbf{SSO} & \textbf{Appearance} & \textbf{\begin{tabular}[c]{@{}c@{}}Functionality \end{tabular}} \\ \midrule
\textbf{Minor} & \highlight[green]{0\%} & \highlight[yellow]{1\%} & \highlight[green]{0\%} & \highlight[yellow]{3\%} \\

\textbf{Major} & \highlight[green]{0\%} & \highlight[red]{11\%} & \highlight[green]{0\%} & \highlight[red]{3\%} \\ \bottomrule
\end{tabular}
\caption{Qualitative manual analysis for 100 websites using \name, showing \% of \highlight[green]{No} , \highlight[yellow]{Minor}, and \highlight[red]{Major} breakages in navigation, SSO, appearance, and functionality.}
\label{tab:web-breakage}
\end{table}

\subsection{Runtime Performance}
\label{subsec:runtime_performance}
We evaluate performance overhead on the top 10K websites by visiting each URL with and without \name and collecting standard page-load metrics via Selenium. 
To ensure a fair comparison, we keep only URLs observed in \emph{both} conditions and discard invalid or non-positive measurements. After pairing and cleaning, our analysis covers 8{,}171 valid websites.

These metrics include the timing of key page load events. 
Specifically, we measure \texttt{dom\_content\_loaded}, \texttt{dom\_interactive}, and \texttt{load\_event\_time} using Selenium.
\texttt{dom\_content\_loaded} time indicates when the HTML document has been completely loaded and parsed, without waiting for stylesheets, images, and subframes to finish loading.
\texttt{dom\_interactive} time indicates the duration until the DOM is fully prepared for user interaction. \texttt{load\_event\_time} indicates the total time to completely load all resources, such as images and CSS, signifying the full usability of a web page. 

Table \ref{tab:web-breakage} summarizes the results. 
A standard benchmark \cite{smith2021sugarcoat} for user-perceived web page performance measures the timeline of key events marking various stages in the browser’s page load and rendering process. 
The management of cookies by \name adds an average overhead of 0.3 seconds to website performance.
While this may impact website load times negligibly, it is a reasonable trade-off for the significant improvements in security and privacy controls that \name provides. 
By managing cookie access more strictly, \name prevents unauthorized data access and enhances user trust.

\begin{table}[!htb]
\centering
\small
\begin{tabular}{@{}lll@{}}
\toprule
\multicolumn{1}{c}{\textbf{Metrics}} &
  \multicolumn{1}{c}{\textbf{\begin{tabular}[c]{@{}c@{}}Normal\\ (Mean, Median)\end{tabular}}} &
  \multicolumn{1}{c}{\textbf{\begin{tabular}[c]{@{}c@{}}\name\\ (Mean, Median)\end{tabular}}} \\ \midrule
\textbf{DOM Content Loaded} & 1659 ms, 946 ms & 1896 ms, 1020 ms \\
\textbf{DOM Interactive}    & 1464 ms, 842 ms & 1702 ms, 911 ms \\
\textbf{Load Event}         & 3197 ms, 2008 ms & 3635 ms, 2136 ms \\ \bottomrule
\end{tabular}
\caption{Performance analysis for the \name replacement}
\label{performance}
\end{table}

\begin{figure*}[t]
  \centering
  \begin{subfigure}{.32\linewidth}
    \includegraphics[width=\linewidth]{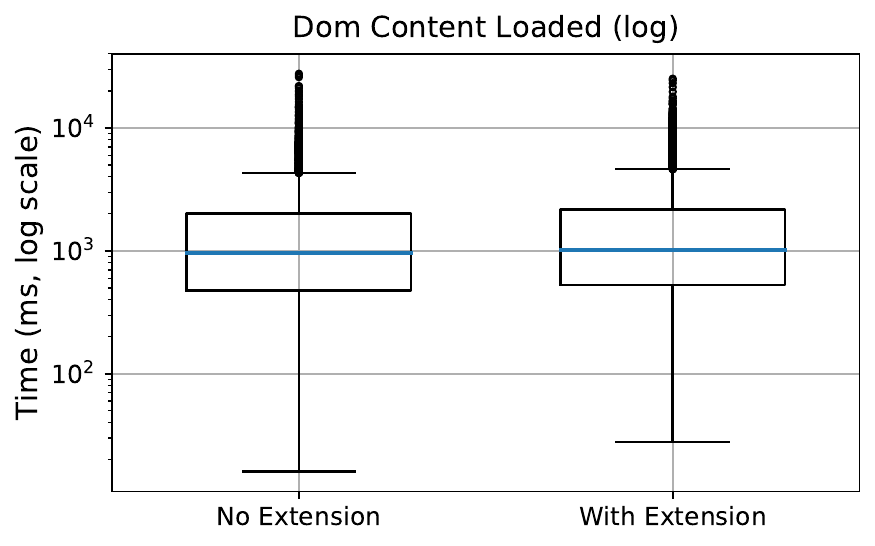}
    \caption{DOM Content Loaded.}
    \label{fig:dcl-box}
  \end{subfigure}\hfill
  \begin{subfigure}{.32\linewidth}
    \includegraphics[width=\linewidth]{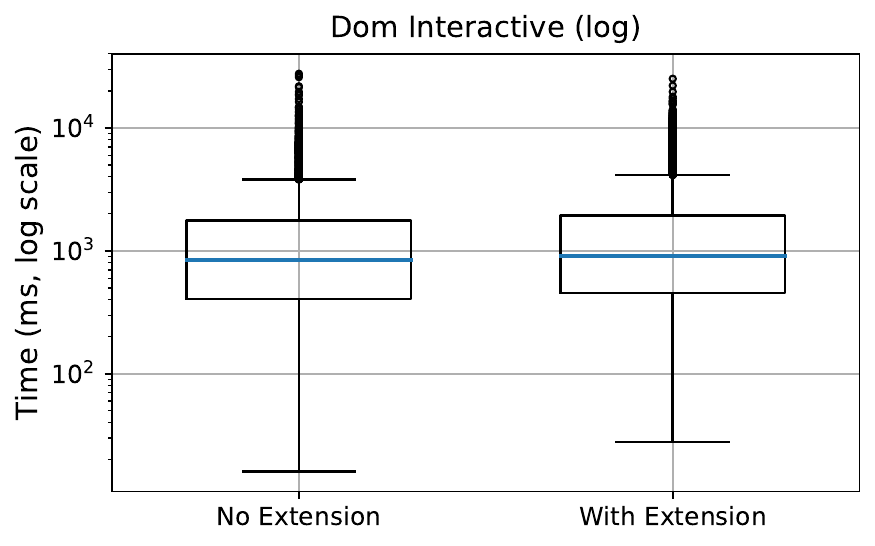}
    \caption{DOM Interactive.}
    \label{fig:di-box}
  \end{subfigure}\hfill
  \begin{subfigure}{.32\linewidth}
    \includegraphics[width=\linewidth]{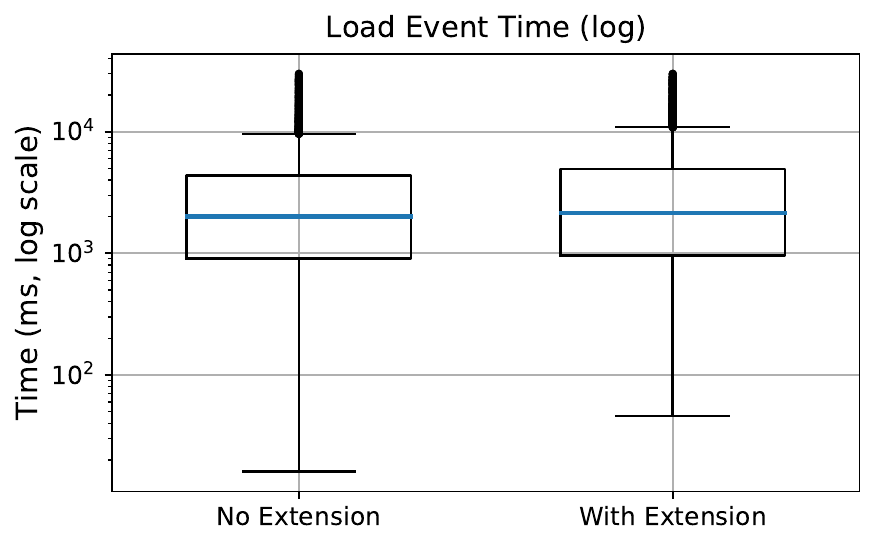}
    \caption{Load Event Time.}
    \label{fig:let-box}
  \end{subfigure}
  \caption{Paired distribution of DOM Content Loaded, DOM Interactive, and Load Event Time (With \name vs.\ No \name) with logarithmic y-axis. Blue lines show medians; boxes show interquartile range (IQR); whiskers indicate 1.5$\times$IQR; points are outliers.}
  \label{fig:runtime-boxplots-log}
\end{figure*}

\textbf{Distributional view.}
Means and medians alone can obscure the heavy-tailed, multiplicative nature of page-load times; a small fraction of very slow loads disproportionately influence aggregates.
Therefore, we visualize \emph{paired} distributions for DOM Content Loaded, DOM Interactive, and Load Event Time on a logarithmic y-axis (Figure ~\ref{fig:runtime-boxplots-log}). 
The two conditions (“No Extension” and “With \name”) appear as separate x-axis categories with identical styling; medians are center lines, boxes show interquartile ranges, whiskers extend to 1.5×IQR, and points denote outliers. 
The log scale makes factor-level differences legible and reveals long right tails typical of modern pages. 
Across all three metrics, the \emph{With \name} boxes are slightly shifted upward, indicating a modest but systematic overhead.
The tail is most pronounced for \textsc{Load Event Time}, which, by definition, waits for all subresources and any \texttt{window.load} handlers; pages with extensive third-party stacks dominate this tail.

\begin{figure}[t]
  \centering
  \includegraphics[width=.75\linewidth]{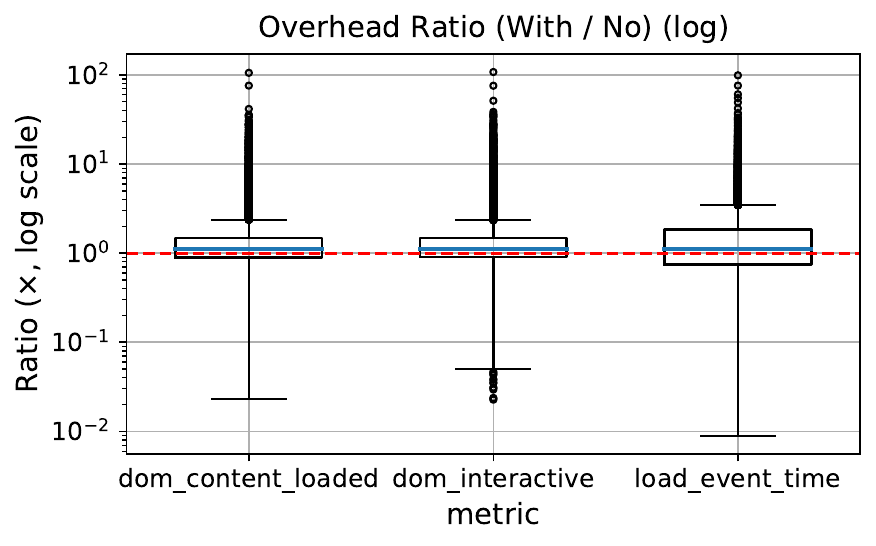}
  \caption{Overhead ratio (With \name\,/\,No \name) for each metric, log-scale y-axis. The dashed line at 1.0 indicates parity.}
  \label{fig:ratio-boxplot-log}
\end{figure}

We also report a per-site overhead ratio (With/No) on a log axis (Figure \ref{fig:ratio-boxplot-log}) to normalize across websites with very different absolute times. The dashed line at 1.0 marks parity; medians above it indicate a typical slowdown while the spread reflects cross-site variability. Over 8,171 paired sites, median ratios are 1.108 for DOM Content Loaded, 1.111 for DOM Interactive, and 1.122 for Load Event Time.
Linear (millisecond) versions of the three boxplots, along with absolute summaries, appear in the Appendix \ref{app:absolute-ms} for direct reading of raw timing differences.

% These improvements are seen across several metrics: DOM content load time, which indicates the time from page load start to complete, HTML parsing, and initial DOM interactive time. 
%
% The decrease in load times and improvement in performance metrics when using the \name extension can be attributed to several factors related to how web resources are handled and restricted. 
% %
% Potential key reasons could be as follows:
% %
% \begin{itemize}
%     \item Fewer HTTP Requests: With restrictions on cookie access by cross-domain scripts, there are generally fewer HTTP requests made. 
%     %
%     For example, cookie syncing is a practice to exfiltrate first-party cookies to third-party domains.
%     %
%     Since \name prevents cross-domain scripts from getting access to cookies they did not set, the number of network requests decreases.
%     %
%     Each HTTP request adds overhead due to connection setup, data transmission, and waiting for the server's response. 
%     %
%     By reducing the number of these requests, particularly those involving third-party domains, the overall page load time can be noticeably decreased.

%     \item Decrease in Background Activity: Many third-party scripts perform tasks in the background (like tracking, advertising, or analytics).
%     %
%     These tasks can consume system resources and slow down the user experience. 
%     %
%     Limiting these scripts' access to cookies can reduce their functionality, thus preserving system resources and enhancing page load times.
% \end{itemize}

\section{Limitations and Future work}
\name's  prototype implementation highlights both the promise and challenges of ownership-based cookie isolation. In this section, we describe its limitations and point to future work that can improve its practicality and expand its applicability.

% \textbf{Blocklist Evasion}
% Traditional defenses often rely on blocklists of known tracker domains. However, blocklists are inherently limited: trackers frequently register new domains or rotate existing ones to bypass filter rules, creating a persistent cat-and-mouse dynamic between trackers and blocklist maintainers. CookieGuard avoids this weakness by design. Rather than relying on enumerating tracker domains, CookieGuard enforces per-script domain isolation of the cookie jar. This ensures that even if a tracker uses a newly minted or randomized domain, it cannot leverage inclusion in the main frame to exfiltrate or overwrite cookies belonging to other domains.

% \textbf{Script Obfuscation}
% Another common evasion tactic is code obfuscation, where trackers disguise or transform their scripts to avoid detection. CookieGuard’s enforcement does not depend on code structure, syntax, or labeling, but solely on execution context (the script’s domain provenance). Because enforcement is context-based, obfuscated scripts are subject to the same restrictions as unobfuscated ones. Stack traces still reveal script URLs regardless of code complexity, meaning CookieGuard applies its isolation policy uniformly, making obfuscation ineffective as an evasion strategy.
% \subsection{Limitations}
\label{section:limitation}
\textbf{Manipulation of script source.}
%A key design choice Section \ref{cookie_guard_design}, first-party scripts have unrestricted access to the cookie jar.
%
\name relies on the script source attribute to distinguish between third-party scripts from different domains in the main frame.
However, these third-party scripts can attempt to manipulate the script source to circumvent \name's protection in the following ways:

\begin{itemize}
    \item Embedded as inline scripts: rather than including scripts using script tags, the source of a third-party script can be embedded as inline script. Inline scripts are automatically considered as first-party scripts.
    \item Hosted as a first-party script: web developers may decide to host the script code on the same domain as the website (or a subdomain of the website).
    \item CNAME cloaking: CNAME cloaking \cite{lin2022investigating, snyder2020filters} is a technique that creates a first-party subdomain and assigns a DNS CNAME record pointing to the third-party domain of the script.
\end{itemize}
As discussed in Section~\ref{label:cookie_guard_design}, \name adopts a safe-by-default strategy for inline scripts by treating them as untrusted and denying access to all cookies.
An alternative approach would be to assess their behavior at runtime and determine trust based on observed execution patterns.
Chen et al. \cite{chen2021detecting} propose a signature-based mechanism that creates behavior signatures for scripts. 
A practical method would involve conducting a large-scale web crawl to generate signatures for various third-party scripts. 
If \name detects that a first-party script's signature matches that of a third-party script, it can then treat it as a third-party script.
A key challenge that warrants further attention is JavaScript obfuscation~\cite{moog2021}. 
For this signature-based approach to work, the signatures should be resistant to minor changes in a script's code.

We acknowledge that CNAME cloaking and intentional self-hosting by site owners are difficult challenges for any client-side defense. However, our focus with \name is primarily on the more common and less deliberate cases where third-party scripts are introduced indirectly, often via tag managers or bundled SDKs, in ways that web developers may not be fully aware of. In these scenarios, our solution provides meaningful protection by preventing unintended cross-script access and exfiltration. When a web developer intentionally colludes with trackers (e.g., through CNAME cloaking) the problem becomes significantly harder, as attribution is intentionally obscured at the network or DNS layer. In such cases, existing DNS-based defenses against CNAME cloaking can also help \name \cite{brave2020cname,wilander2020cname,nextdns2025cname}.

\textbf{Toward Practical Deployment.}
Currently, \name's prototype implementation is available to use as a browser extension rather than a ready-to-deploy browser feature. In practice, browser vendors have historically introduced potentially disruptive privacy protections incrementally, giving developers time to adapt. For example, Safari rolled out Intelligent Tracking Prevention (ITP) in stages—starting in 2017 with limits on cross-site cookies and link-decoration controls \cite{ITP}, then moving to full third-party cookie blocking in March 2020 \cite{ITP2.3}. During this transition, WebKit employed temporary “grandfathering” mechanisms to preserve existing site data while ITP relearned user interactions, easing the migration for websites \cite{grandfathering}. A similar path could be envisioned for \name: initially offered as an opt-in feature or in private browsing mode, then incrementally moved toward default enforcement with advance notice to developers. Another possibility would be to expose \name's policies as user-selectable privacy settings, allowing users to balance functionality and privacy according to their preferences.

\textbf{Interactive mode and scope limitations.} Our automated large-scale crawl includes light user interaction to mimic realistic browsing behavior: the crawler scrolls through pages and clicks on three random links per site. 
However, we do not perform any authenticated logins during the crawl, and as a result, authentication cookies are not generated.
Most session cookies used for authentication are correctly marked with the \texttt{HttpOnly} flag, which makes them inaccessible to JavaScript and, thus, out of scope for the class of script-based attacks we study. 
As described in Section~\ref{sec:background_cookie_management}, browser vendors enforce this restriction specifically to protect sensitive cookies from exfiltration via client-side scripts.

Furthermore, because our crawler does not fill forms or input personal data, we did not observe the leakage of personally identifiable information (PII) during cookie exfiltration. 

Exploring script-based exfiltration of authentication tokens or PII in authenticated browsing contexts remains an important direction for future work. 
We plan to complement our current analysis with user-assisted or session-replay-based authenticated crawls to investigate these higher-risk scenarios.

\textbf{Limitations of Source Attribution.}
\name relies on the JavaScript stacktrace to identify the script origin responsible for cookie access. 
This approach is effective in most synchronous and directly-invoked contexts, including event listeners and DOM-triggered handlers. 
However, it may fall short in certain asynchronous scenarios—such as when cookies are accessed in callbacks following \texttt{setTimeout}, \texttt{Promise} resolutions, or event queue re-entries—where the originating script may no longer appear in the stacktrace. 
While recent browser APIs like \texttt{Error.prepareStackTrace} and \texttt{async stack traces} can improve coverage, some edge cases remain unresolved.
% \name also does not currently intercept cookie access from scripts executing in isolated contexts such as \texttt{srcdoc} iframes or \texttt{blob:} URLs. 
% %
% These scripts execute in their own execution environments, and their connection to the originating third-party script is not always preserved or exposed by the browser. 
% %
% Handling such cases would require broader instrumentation of the browser's frame and blob URL creation logic—potentially at the browser engine level rather than within an extension. 
%
We leave a more comprehensive treatment of such indirect flows to future work.

% \textbf{Attribution via Stack Traces.}
% \edit{Our method for attributing cookie accesses relies on stack traces, which in practice provide reliable provenance for main-frame scripts—the scope of our study. Script obfuscation does not undermine this approach, as stack traces reveal the script URL regardless of code complexity. The main limitation arises with asynchronous callbacks, where the original script may fall off the stack. We acknowledge that our methodology may conservatively undercount some asynchronous accesses.}

\textbf{Functional Motivations for Cross-Site Cookie Access.}
Our manual evaluation shows that most breakage under \name is due to Single Sign-On (SSO) systems, where identity providers (e.g., Google, Microsoft, Okta) access or modify first-party cookies to manage login flows. Blocking such access can disrupt session continuity.

Other legitimate use cases—such as analytics, A/B testing, or personalization—may also involve cross-domain access to first-party cookies. However, these are often indistinguishable from privacy-invasive practices like tracking or identifier syncing.

\name adopts a strict isolation model by default but can be extended to support whitelisting for trusted domains when needed.

% \subsection{Future Work}

\textbf{Beyond cookies: Cross-domain manipulation of other shared resources.}
As we discussed in Section \ref{section:threat_model}, scripts included in the main frame have access to a variety of shared resources beyond cookies such as the HTML DOM.
We conducted a pilot analysis to investigate the existence and prevalence of cross-domain DOM modification.
We observed that a number of cross-domain scripts run with full privileges modify, insert, or remove DOM elements that do not belong to them on 9.4\% of sites. 
Modification can be applied to the content of HTML elements, e.g., \textit{innerText} or \textit{innerHTML}, the style (CSS) of the HTML element, and the attributes and classes of the HTML element, e.g., \textit{src}. 
We leave a more comprehensive investigation of this issue to future work and plan to develop a targeted defense mechanism to mitigate this behavior.

% %
% %

% \xxx[Umar]{This is super cool! I really like this idea and would like to see it expanded beyond cookies and towards characterizing all interactions between scripts and other resources loaded in the same origins. Then also exploring a more comprehensive solutions for moderating all such interactions!}

\section{Related Work}
\label{sec: related work}

\textbf{Third-party scripts.}
Prior research has studied the widespread inclusion of third-party scripts on websites and security risks associated with their use \cite{nikiforakis2012you, lauinger2018thou, musch2019scriptprotect}.
Lauinger et al. \cite{lauinger2018thou} have investigated over 133K websites finding that 37\% of them include at least one script with a known vulnerability.
Musch et al. \cite{musch2019scriptprotect} modify the JavaScript environment in a such a way that the accidental introduction of a Client-Side XSS vulnerability through a third party is prevented.
Nikiforakis et al. \cite{nikiforakis2012you} analyze the extensive presence of third-party scripts across over 3 million pages from the top 10,000 Alexa sites, reporting that 88.5\% of popular sites incorporate at least one third-party script, not necessarily in the main frame. 
The study also tracks the increasing reliance on third-party scripts over time.
In contrast, our research, unlike others that focus on security, explores the privacy implications of third-party scripts in the main frame.
Additionally, our study specifically targets the inclusion of third-party scripts in the main frame, finding that 93.3\% of websites include at least one such script directly in their main frame. 
Our analysis categorizes these scripts by their method of inclusion, revealing that only 10\% are directly included, while a significant 90\% are included indirectly across 10,000 websites. 
This suggests that the rise in third-party script inclusion may not directly correlate with increased trust by websites.

% web alamnac: 
% - https://almanac.httparchive.org/en/2022/javascript#first-party-versus-third-party-javascript

\textbf{First-party cookies ghost-written by third-party scripts.}
Another line of prior work has studied cookie creating and exfiltration by third-party scripts. 
Sanchez et al. \cite{sanchez2021journey} analyzed cookie (both script and HTTP cookies) exfiltration, overwriting, and deleting across 1 million websites.
They collected 66.7 million cookies from 74\% (738,168) of the websites visited. 
Their findings reveal that 11\% of these cookies are first-party, 47\% are third-party, and 42\% are ghost-written. 
They also find that 13.4\% of all cookies are exfiltrated, 0.19\% are overwritten, and 0.08\% are deleted by scripts or Set-Cookie headers in responses. 
Additionally, they report that cross-domain cookie exfiltration and collision (including overwriting and deleting) occur on 28.3\% and 0.7\% of the visited websites, respectively.
Our work focuses exclusively on first-party scripts (created by both first-party and third-party scripts on the main frame) due to their critical role in storing users' sensitive information and session cookies. 
Consequently, we only consider cross-domain actions, including exfiltration and manipulation, by scripts on the main frame. 
We do not investigate actions caused by Set-Cookie headers in responses, as responses sent from third-party servers cannot access or manipulate first-party cookies.
Since the publication of this paper, i.e. 2021, we have noted a significant shift of third-party scripts into the main frame. 
Notably, 92\% of all first-party cookies in our dataset are ghost-written cookies.
This shift has had significant consequences. 
The incidence of cross-domain cookie exfiltration, overwriting, and deletion has increased by factors of 1.3, 19.5, and 6.2, respectively. 
Additionally, the number of websites engaging in cross-domain cookie exfiltration, overwriting, and deleting has risen by 1.5 and 45.7 times, respectively. 
It is important to note that these figures represent a conservative estimate, as our study focuses only on first-party and ghost-written cookies manipulated or exfiltrated by scripts in the main frame, excluding other scripts and HTTP responses.

Chen et al. \cite{chen2021cookieswap} investigated ghost-written first-party cookies set by third-party tracking scripts. 
Their findings indicate that 97\% of the Alexa top 10K websites host at least one ghost-written cookie, with exfiltration occurring on 57.7\% of these sites by both same-domain and cross-domain scripts. 
However, their study does not address other impacts of third-party scripts in the main frame, such as cookie manipulation. 
Moreover, neither the studies by \cite{sanchez2021journey} nor \cite{chen2021cookieswap} offer a mechanism to protect first-party cookies from cross-domain exfiltration and manipulation.
There are extensive number of tools and extensions allowing users to automatically delete cookies from specified domains \cite{AutoDelete, CookieQuickManager, uMatrix}.
Moreover, Munir et al.\cite{munir2022cookiegraph} proposed a machine learning-based mechanism to automatically block ghost-written first-party tracking cookies. 
However, none of these studies address the more fundamental problem of first-party cookie isolation, i.e., whether first-party cookies can be accessed or
modified by different third-party scripts.

\textbf{Shared global namespace.}
Another line of prior work has studied JavaScript conflicts in the shared global namespace.
Ocariza et al. \cite{ocariza2016study} and Zhang et al. \cite{zhang2020detecting} investigated how JavaScript global identifier conflicts might lead to cookie overwriting. 
Several studies focused on JavaScript issues, predominantly focusing on bad coding practices \cite{ocariza2012autoflox, ocariza2016study, hackett2012fast, jensen2009type, pradel2015typedevil, zhang2020detecting, patra2018conflictjs}. 
The primary objective of these studies is to identify conflicts shared global namespace, especially when they result in functional web issues, as opposed to focusing solely on shared cookie jar on the main frame.

\textbf{Takeaway.} 
In summary, our work is distinct from prior work in two main ways.
First, our study focuses on first-party cookie manipulation/exfiltration by third-party scripts in the \textit{main frame}.
Second, we propose a defense mechanism called \name to isolate cookies from manipulation/exfiltration by cross-domain scripts.

\section{Conclusion}
In this work, we conducted a large-scale measurement of cross-domain cookie access by third-party scripts in the main frame and introduced \name, a browser-based system to isolate first-party cookies.
\name advances the state-of-the-art in two major ways.
First, different from prior work on third-party JavaScript and cookie measurements, we specifically focus on third-party scripts included -- either directly or indirectly -- in the main frame such that they can manipulate/exfiltrate cookies in the first-party cookie jar.
Second, different from prior work on blocking or partitioning third-party and/or first-party cookies, \name is the first-of-its-kind first-party cookie partitioning system.
\name complements interventions that are being introduced in web browsers to improve the security and privacy of cookies. 
Going forward, the underlying problem addressed by \name goes well beyond cookies. 
Third-party scripts included in the main frame have access to \textit{all} first-party shared resources (e.g., HTML DOM, global namespace), not just the cookie jar. 
Future work can develop interventions, similar to \name for the cookie jar, to introduce isolation in other shared resources.

% Note from the CFP that this section must include a statement about
% ethical issues; papers that do not include such a statement may be
% rejected.

%%%%%%%%%%%%%%%%%%%%%%%%%%%%%%%%%%%%%%%%%%%%%%%%%%%%%%%%%%%%%%%%%%%%%%%%%%%%
% We're in the endgame now

\bibliographystyle{ACM-Reference-Format}
\bibliography{main}

\appendix

\section{Appendix}

\subsection{Frequently overwritten or deleted cookies}
\label{appendix:top_manipulator}
Table \ref{tab:manipulation-cookie-attacks} presents the frequently manipulated cookie pairs by entities across 20,000 websites.

\begin{table*}[!htb]
\centering
\scriptsize
\begin{adjustbox}{width=\textwidth}
\begin{tabularx}{\textwidth}{l|l l r X}
\toprule
\makecell{\textbf{Manipulation}\\\textbf{Type}} & 
\makecell{\textbf{Cookie}\\\textbf{Name}} & 
\makecell{\textbf{Creator}\\\textbf{Domain}} & 
\makecell{\textbf{Number of Manipulator}\\\textbf{Entities}} & 
\makecell{\textbf{Top 3 Manipulator}\\\textbf{Entities}} \\
\midrule
\multirow{10}{*}{Overwriting} 
& \_fbp & facebook.net & 132 & Functional Software, Google, Segment.io \\

& OptanonConsent & cookielaw.org & 132 & Google, New Relic, WarnerMedia \\

& \_ga & googletagmanager.com & 86 & Gatehouse Media, Intergi Entertainment, Segment.io \\

& \_gcl\_au & googletagmanager.com & 49 & playbetter.com, Cybot ApS, bombayshirts.com \\

& \_uetvid & bing.com & 48 & Segment.io, Tealium, forseasky.com \\

& \_uetsid & bing.com & 45 & Segment.io, Tealium, forseasky.com \\

& ajs\_anonymous\_id & segment.com & 45 & Functional Software, Journal Publishing, conwaydailysun.com \\

& cto\_bundle & criteo.com & 39 & Future Plc, script.ac, CacheNetworks \\

& \_gid & google-analytics.com & 34 & Olark, Functional Software, Intergi Entertainment\\

& utag\_main & tiqcdn.com & 33 & UNIDAD Editorial SA, medica.com, Optimizely \\

\midrule
\multirow{10}{*}{Deleting}
& \_uetvid & bing.com & 54 & Tealium, Segment.io, cookie-script.com \\

& \_uetsid & bing.com & 51 & Tealium, Segment.io, cookie-script.com \\

& \_ga & googletagmanager.com & 27 & cdn-cookieyes.com, Civic Computing, cookie-script.com \\

& \_fbp & facebook.net & 18 & cdn-cookieyes.com, cookie-script.com, Civic Computing \\

& \_gid & google-analytics.com & 18 & cdn-cookieyes.com, cookie-script.com, elfsborg.com \\

& \_gcl\_au & googletagmanager.com & 18 & cdn-cookieyes.com, cookie-script.com, Civic Computing \\

& \_cookie\_test & cxense.com & 15 & Enreach, optable.co, Canadian \\

& \_ga & google-analytics.com & 14 & cdn-cookieyes.com, cookie-script.com, elfsborg.com\\

& ajs\_user\_id & segment.com & 13 & Functional Software, solvhealth.com, printify.com \\

& \_screload & snapchat.com & 11 & Cisco, sparksites.io, acdc.com \\
\bottomrule
\end{tabularx}
\end{adjustbox}
\caption{Frequently overwritten and deleted cookies by other entities}
\label{tab:manipulation-cookie-attacks}
\end{table*}

\section{Top cross-domain manipulators}
\label{appendix:top_manipulator_domains}

\begin{figure*}[t]
\centering
\begin{subfigure}[t]{0.49\textwidth}
    \centering
    \includegraphics[width=0.8\textwidth]{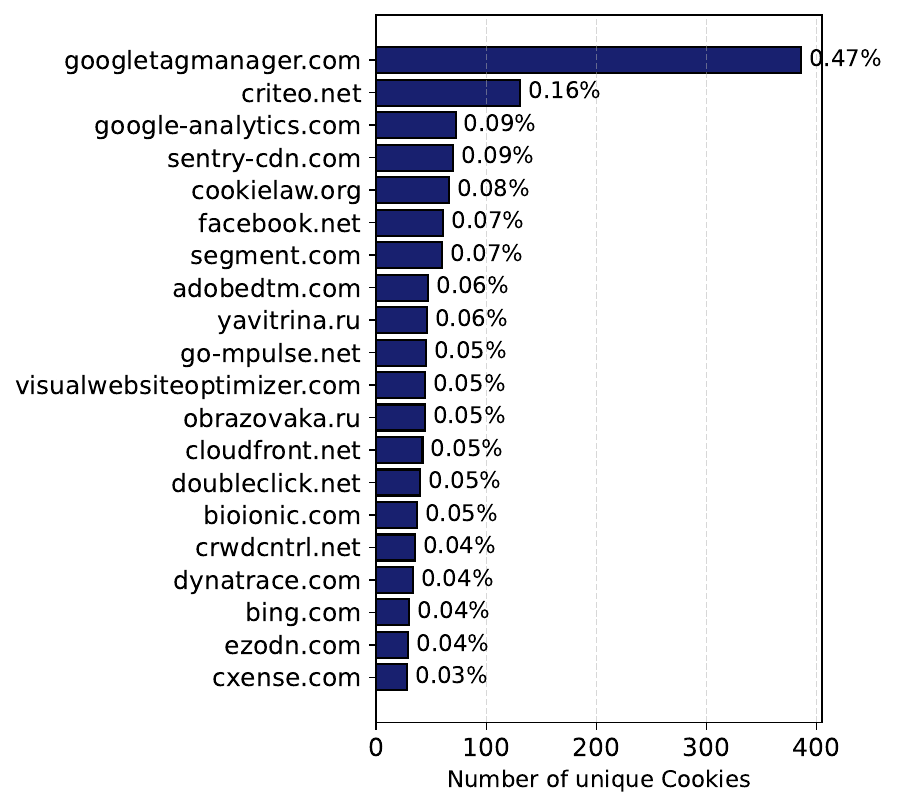}
    \caption{Domains engaged in cross-domain cookie overwriting}
    \label{fig:top_overwriter}
\end{subfigure}
\hfill
\begin{subfigure}[t]{0.49\textwidth}
    \centering
    \includegraphics[width=0.8\textwidth]{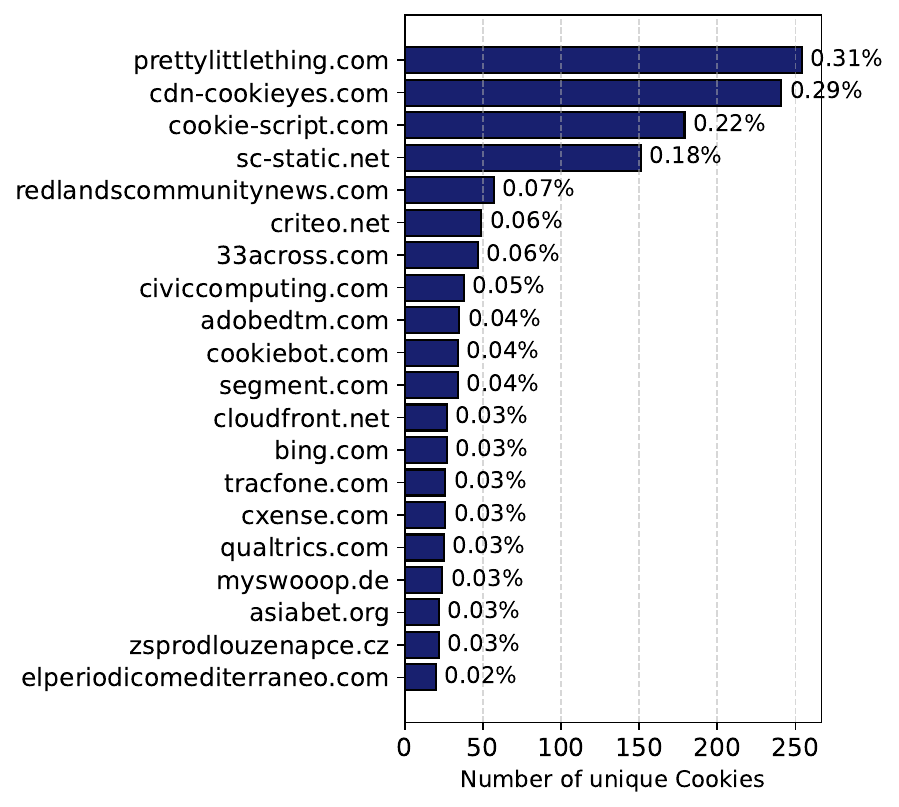}
    \caption{Domains engaged in cross-domain cookie deleting}
    \label{fig:figure2}
\end{subfigure}
\caption{Top 20 script-hosting domains engaged in cross-domain overwriting and deleting unique cookies across 20k websites ranked by ranked by the number of cookies they manipulate.}
\label{fig:top_manipulator_domains}
\end{figure*}

Figure \ref{fig:top_manipulator_domains} presents top  Top 20 script-hosting domains engaged in cross-domain overwriting and deleting.

\section{Absolute Distributions (Linear Scale)}
\label{app:absolute-ms}
\begin{figure*}[t]
  \centering
  \begin{subfigure}{.32\linewidth}
    \includegraphics[width=\linewidth]{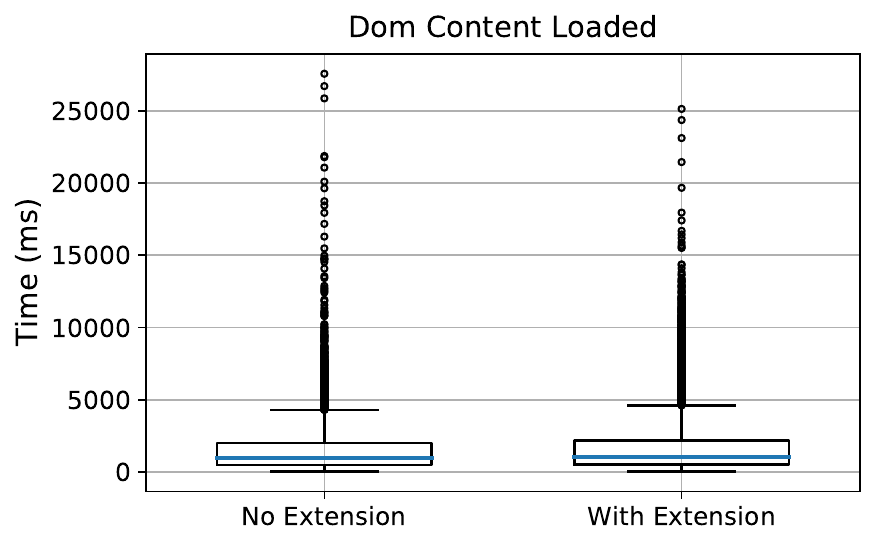}
    \caption{DOM Content Loaded.}
  \end{subfigure}\hfill
  \begin{subfigure}{.32\linewidth}
    \includegraphics[width=\linewidth]{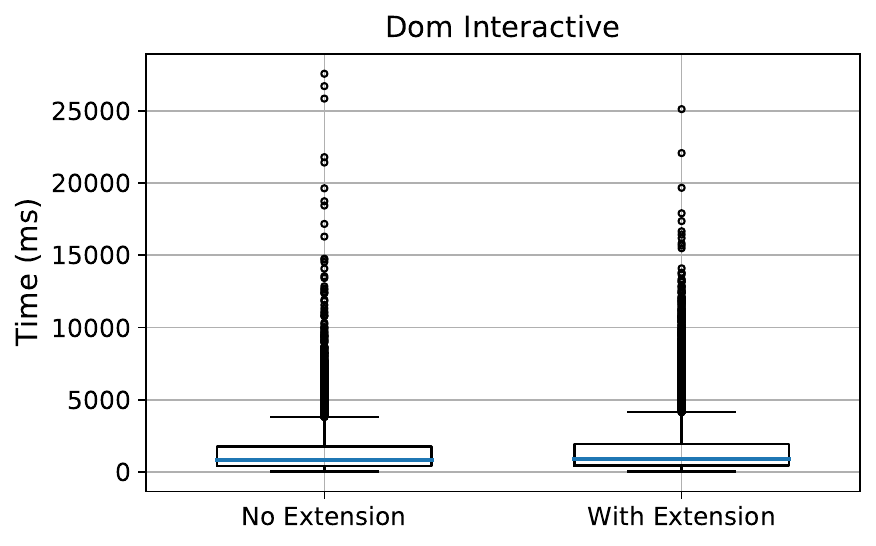}
    \caption{DOM Interactive.}
  \end{subfigure}\hfill
  \begin{subfigure}{.32\linewidth}
    \includegraphics[width=\linewidth]{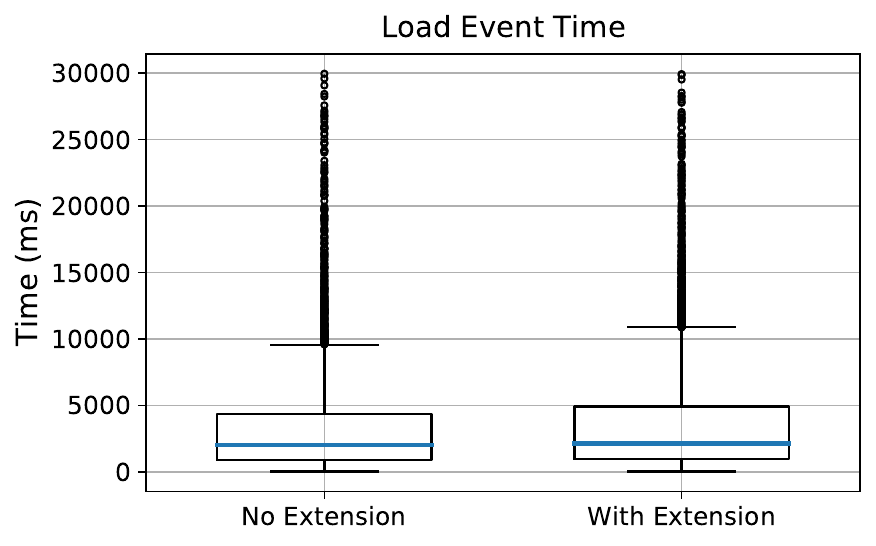}
    \caption{Load Event Time.}
  \end{subfigure}
  \caption{Paired boxplots in milliseconds (No Extension vs.\ With Extension).}
  \label{fig:runtime-boxplots-linear}
\end{figure*}

To complement the log-scale views in Section \ref{subsec:runtime_performance}, we plot the same paired distributions on a \emph{linear} y-axis (milliseconds) in Figure ~\ref{fig:runtime-boxplots-linear}. 
% As before, categories on the x-axis indicate the two conditions (No Extension vs.\ With Extension), and both use identical styling. 
The linear scale highlights absolute timing while making clear how a small number of very slow pages can extend the whiskers/outliers.

\begin{figure*}[]
  \centering
  \includegraphics[width=.5\linewidth]{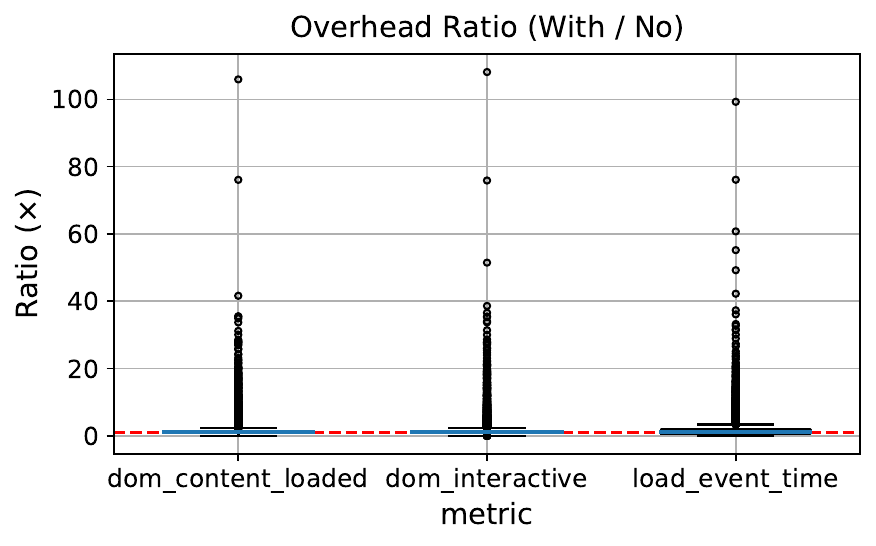}
  \caption{Per-site overhead ratio (With \name\,/\,No \name), linear y-axis.}
  \label{fig:ratio-boxplot-linear}
\end{figure*}

We also include the per-site \emph{overhead ratio} (With/No) on a \emph{linear} y-axis in Figure ~\ref{fig:ratio-boxplot-linear}. 
Whereas the log-scale ratio in Figure \ref{fig:ratio-boxplot-log} makes multiplicative tails visually comparable, the linear view emphasizes deviations near parity (1.0): small slowdowns or speedups are easier to read, while very large ratios are visually compressed.

% On a linear axis, vertical differences can be read directly in milliseconds (absolute effect). The central boxes indicate typical ranges; extended whiskers and many outliers reflect heavy pages with large resource sets or expensive \texttt{window.load} handlers (especially for Load Event Time). These absolute views corroborate the log-scale takeaway: a modest, consistent overhead on typical pages and a long right tail on a minority of sites.

\section{Ethics}
This study adheres to established ethical guidelines for research involving data collection and analysis. Our methodology is based solely on the examination of publicly available data retrieved from websites. Specifically, we analyze cookies stored on our own machines and outbound network requests initiated during automated browsing. At no point did our research involve the collection, interception, or processing of any real user's Personally Identifiable Information (PII). We exclusively analyzed publicly accessible scripts and configuration data as served by websites. Therefore, we believe our work does not raise any ethical concerns.

\end{document}